\documentclass[journal,twoside,web]{ieeecolor}
\usepackage{generic}
\usepackage{cite}
\usepackage{amsmath,amssymb,amsfonts}

\usepackage{algorithmic}
\usepackage{graphicx}
\usepackage{textcomp}

\usepackage{lipsum}

\usepackage{subfigure}
\usepackage{comment,color}

\usepackage{subfigmat}
\usepackage{caption}

\newcommand{\R}{\ensuremath{{\mathbb R}}}

\newcommand{\N}{\ensuremath{{\mathbb N}}}

\newcommand{\NN}{{\mathcal N}}
\newcommand{\OO}{{\mathcal O}}
\newcommand{\ra}{\rightarrow}

\newcommand{\AAA}{\mathsf{A}}
\newcommand{\CCC}{\mathsf{C}}

\newcommand{\col}{\mathrm{col}}

\newcommand{\xxx}{\mathsf x}

\newcommand{\XXXX}{\mathcal X}

\newtheorem{thm1}{\bf Theorem}
\newtheorem{prop1}{\bf Proposition}
\newtheorem{lem1}{\bf Lemma}
\newtheorem{assmpt1}{\bf Assumption}
\newtheorem{defn1}{\bf Definition}
\newtheorem{rem1}{\bf Remark}

\newtheorem{cor1}{\bf Corollary}

\newtheorem{prob1}{\bf Problem}

\newenvironment{defn}{\begin{defn1}}{\hfill$\square$\end{defn1}}
\newenvironment{asm}{\begin{assmpt1}}{\hfill$\square$\end{assmpt1}}
\newenvironment{rem}{\begin{rem1}}{\hfill$\square$\end{rem1}}

\newenvironment{cor}{\begin{cor1}}{\hfill$\square$\end{cor1}}
\newenvironment{prop}{\begin{prop1}}{\hfill$\square$\end{prop1}}
\newenvironment{prob}{\begin{prob1}}{\hfill$\square$\end{prob1}}

\def\BibTeX{{\rm B\kern-.05em{\sc i\kern-.025em b}\kern-.08em
    T\kern-.1667em\lower.7ex\hbox{E}\kern-.125emX}}
\begin{document}
\title{
\huge
Complexity Reduction for Resilient State Estimation\\ of Uniformly Observable Nonlinear Systems 
}
\author{Junsoo Kim, \IEEEmembership{Member, IEEE},
Jin Gyu Lee, \IEEEmembership{Member, IEEE},
Henrik Sandberg, \IEEEmembership{Fellow, IEEE},\\
and
Karl H. Johansson, \IEEEmembership{Fellow, IEEE}
\vspace{-5mm}
\thanks{
This work was supported
in part by
Seoul National University of Science \& Technology,
	and in part by the Swedish Research Council.
}
\thanks{
J.~Kim is with
Department of Electrical and Information Engineering, Seoul National University of Science and Technology, Korea.
}
\thanks{
H.~Sandberg and K.\,H.~Johansson are with the Division of Decision and Control System, KTH Royal Institute of Technology, Sweden.
They are also affiliated with Digital Futures.
}
\thanks{
J.\,G.~Lee is with
Inria, University of Lille, CNRS, UMR 9189 - CRIStAL, F-59000 Lille, France.
}
}

\maketitle

\begin{abstract}
A resilient state estimation scheme for uniformly observable nonlinear systems, based on a method for local identification of sensor attacks, is presented. The estimation problem is combinatorial in nature, and so many methods require substantial computational and storage resources as the number of sensors increases. To reduce the complexity, the proposed method performs the attack identification with local subsets of the measurements, not with the set of all measurements. A condition for nonlinear attack identification is introduced as a relaxed version of existing redundant observability condition. It is shown that an attack identification can be performed even when the state cannot be recovered from the measurements. As a result, although a portion of measurements are compromised, they can be locally identified and excluded from the state estimation, and thus the true state can be recovered. Simulation results demonstrate the effectiveness of the proposed scheme.
\end{abstract}


\section{Introduction}\label{sec:intro}

In recent years, we have witnessed tremendous advances in network communication and computational power,
which has made control systems more connected and capable.
Networked control has enabled significant development in a number of industrial fields, such as robotics, smart grids, and autonomous automobile systems.

Meanwhile,
while more network connectivity leads to more abilities and versatility,
the network layer is inherently more exposed to unauthorized access from third parties, such as malicious cyber-attacks \cite{Sandberg15,Teixeira15}.
Several such vulnerabilities
have been reported, including the StuxNet worm \cite{Langner11}
and security breaches in water sewage systems \cite{Slay07}.
With the increasing threat of possible cyber-attacks,
a number of studies on security issues in networked systems have been presented \cite{Amin09,Sundaram11,Basar15}.

Detection of cyber-attacks may be fundamentally impossible \cite{Pasqualetti13TAC}, and is more challenging than fault detection.
Typically,
if the majority of the sensors are compromised,
the attack can be undetectable, while the effect of the attack is still disruptive.
For the case when a minority of sensor measurements are corrupted,
the problem of resilient state estimation has been formulated,
which is to
exclude the corrupted measurements and
perform state estimation correctly.
Many results have been presented to
address the problem and handle the attack vulnerability
\cite{Pasqualetti13TAC,Chong15ACC,Kim19TAC,Fawzi14,Shoukry16TAC,Chanhwa19TAC,Jeong21TAC,Junsoo18,Mitra16CDC,Pasqualetti15,Chen18TAC,Lee20TAC,Mao22TAC,Chong20CDC,Shou15CDC,jeon2016resilient}, which is categorized in Section~\ref{subsec:related}.

The methods identify the set of un-corrupted measurement from which the state of the system can be restored.
Assuming that only a limited number of sensors are compromised (no matter how the combination is chosen and how they are changed), 
the corrupted portion is excluded from the state estimation so that
the cyber-attack through the network can not harm the control performance.
The solutions can be used for counteracting, for example,
false data injection to power grids \cite{Liu11} or sensor spoofing attacks \cite{shoukry13springer}.

\subsection{Related Work}\label{subsec:related}

To solve the resilient state estimation problem,
there have been techniques developed based on fault detection and diagnosis \cite{Frank90,Ding08}, as in \cite{Pasqualetti13TAC,Pasqualetti15};
methods motivated from compressed sensing \cite{Tao05,Donoho06}, as in \cite{Fawzi14};
and observer-based methods as in \cite{Chong15ACC,Chanhwa19TAC,Jeong21TAC,Shoukry16TAC,jeon2016resilient}.
Extensions have been presented for decentralized design and distributed operation \cite{Junsoo18,Mitra16CDC,Chen18TAC,Lee20TAC},
nonlinear systems \cite{Kim19TAC,Chong20CDC,Shou15CDC},
and complexity analysis \cite{Mao22TAC}.

One of the main challenges for resilient state estimation is that 
many of the proposed algorithms are combinatorial in nature \cite{Pasqualetti13TAC}. Hence, the algorithms cannot solve the problem in polynomial time in general,
and substantial amount of computational or storage resources may be required,
especially when the number of sensors is large.
For instance,
to identify and isolate/exclude $q$ unknown compromised sensors out of $p$ sensors $(q<p)$,
methods in \cite{Pasqualetti13TAC,Chong15ACC}
construct not less than $\binom{p}{q}$ observers for the computation,
and
methods in \cite{Chanhwa19TAC,Jeong21TAC}
consider $\binom{p}{q}$ cases to search out the un-compromised sensors.

Reducing the complexity of resilient state estimation consequently has been major interest.
Many sufficient conditions that allow complexity reduction have been found,
specifically for linear systems.
For example,
a condition for converting the problem into a convex optimization problem is introduced in \cite{Fawzi14},
and an assumption for making use of gradient decent algorithms is found in \cite{Shoukry16TAC}.
Conditions for dividing
the problem into smaller problems, by a
divide and conquer approach, have been investigated in \cite{Pasqualetti15,Junsoo18}.
Linear systems cases for which the problem is not NP-hard are discussed in \cite{Mitra16CDC,jeon2016resilient,Chen18TAC,Lee20TAC,Mao22TAC}.
Distributed solutions to make the implementation scalable are further developed in \cite{Chen18TAC,Lee20TAC,Mitra16CDC}.

In spite of that most control systems are nonlinear in practice,
reducing the complexity of nonlinear resilient state estimation has rarely been considered.
Exceptions include
the results in \cite{Kim19TAC} and \cite{Chong20CDC}, which extend the approaches of \cite{Chanhwa19TAC} and \cite{Chong15ACC} to a class of nonlinear systems, respectively,
but the order of the computation/storage complexity is at least $\binom{p}{q}$.
Another attempt is made in \cite{Shou15CDC},
in which satisfiability modulo theory is used to reduce the combinatorial complexity,
but the technique
assumes not only that the state but also the input of the system is reconstructed from the outputs,
so that, for example, it is not applicable for systems having non-trivial zero dynamics.

\subsection{Contribution}

We present resilient state estimation for uniformly observable systems,
based on a novel local attack identification method for reducing the computational complexity required.

We first introduce a redundancy notion for nonlinear
sensor measurements,
under which sensor attack identification is possible.
Conventionally,
it has been known that
redundant observability is a (necessary) condition for resilient state estimation (for example, see \cite{Fawzi14} for linear systems, and \cite{Kim19TAC} for uniformly observable systems).
Then,
considering that resilient state estimation is to first identify the un-compromised sensors and then reconstruct the correct state estimate,
we propose that observability itself is not necessary when performing attack identification.
As a result,
the introduced redundancy condition
can be seen as a relaxed version of the existing condition on redundant observability.

Based on the introduced notion of redundant functions,
we propose an attack inspection method,
which guarantees that the presence of attack is detected whenever its size is large enough to be distinguished from measurement noise.
The proposed method checks whether a set of measurement data is included in a predefined set or not, by measuring the distance from the set,
and ensures that the data under inspection is not corrupted if and only if the measured distance is less than a pre-computed threshold.
The threshold is computed from Lipschitz constants of certain left inverse functions related to the redundant functions.
We provide a criterion for checking the Lipschitz property of the left-inverses,
considering practical implementation.
In addition,
we also show how to measure the distance for the nonlinear attack inspection, using a set of sampled data from the underlying state space.

We provide a sufficient condition under which the nonlinear resilient state estimation problem can be divided into smaller local problems and then be solved using only local information and measurements.
The complexity can be significantly reduced when each local problem involves a proper subset of measurements.
Projection mappings are used for
removing a portion of measurement data which are not redundant.

Finally, we provide a class of nonlinear systems for which the proposed local identification method can be implemented in a particularly efficient and constructive way,
including how to find each subset of sensor measurements for the local attack identification, and how to find the projection mappings.
It is shown that all linear systems belong to this class.

\subsection{Organization}

The organization of the rest of this paper is as follows.
Section~\ref{sec:preliminaries} gives the problem formulation and preliminaries.
The problem of resilient state estimation is formally stated,
and by use of high gain observer design with respect to individual sensor measurements,
the problem is represented by a nonlinear equation with presence of corrupted components.
The redundancy notion for the sensor attack identification is introduced in Section~\ref{subsec:redundancy}.
Equivalent conditions implying the ability of attack detection and identification are characterized,
and then the attack inspection method is proposed,
together with guidelines for implementation.
Section~\ref{subsec:estimation} connects the introduced notion to resilient state estimation, and proposes the local identification scheme.
Its benefit in terms of computational effort is discussed.
Then,
Section~\ref{subsec:cases} introduces the case for which the proposed method can be implemented in a more constructive manner,
and show that it is applicable to any linear system.
Finally,
Section~\ref{sec:simulation} illustrates the method on a numerical example, and Section~\ref{sec:conclu} concludes the paper.

\section{Problem Formulation and Preliminaries}\label{sec:preliminaries}

\subsection{Notation}
The set of natural numbers and
real numbers are denoted by $\N$ and $\R$, respectively.
For $p\in\N$, we define $[p]:= \{1,2,\dots,p\}$.
For a sequence $v_1,\dots,v_p$ of column vectors or scalars,
we define $\col\{v_i\}_{i=1}^{p}:=[v_1^\top,\dots,v_p^\top]^\top$.
For a sequence $\NN=(n_1,n_2,\dots,n_p)$ of natural numbers,
we define $\R^\NN$ as
the Cartesian product of Euclidean spaces, as
$$\R^\NN:= \R^{n_1}\times\R^{n_2}\times\dots\times\R^{n_p}.$$
For a vector $v=
\col\{v_i\}_{i=1}^{p}
\in\R^\NN$,
the number of non-zero components of $v$ is denoted as
$\|v\|_0^\NN= \left|\{i\in [p]: v_i\neq 0\} \right|,$
where $\left|\cdot\right|$ is the cardinality of a set.
For $v=\col\{v_i\}_{i=1}^{p}\in\R^\NN$,
the canonical projection
with respect to the index set $ I=\{i_1<i_2<\dots< i_l\}\subset[p]$
is denoted as
$\pi_I(v):=
\col\{v_{i_j}\}_{j=1}^{l}
$,
and we write $v_I:=\pi_I(v)$. 
For a function $\Phi: X \ra \R^\NN$, $\left| \NN \right|=p$, and $ I\subset[p]$, the composition of $\pi_I$ and $\Phi$ is denoted as $\Phi_I:=\pi_I\circ\Phi$.
The infinity norm of a vector $v\in\R^n$ is denoted by $\| v\|$.
For a compact set $\XXXX$ and a vector $z$ in $\R^n$,
we define
$\|\XXXX\|:= \max_{x\in\XXXX}\|x\|$ and $d(z,\XXXX):=\min_{x\in\XXXX}\|z-x\|$.
A function $\phi$ defined on a set $X$ is called Lipschitz (on $X$),
if
there exists a constant $L$ such that
$\|\phi(x_1)-\phi(x_2) \|\le  L\|x_1-x_2\|$ holds,
$\forall x_1,x_2\in X$.
We let $L(\phi)$ denote the infimum of such constant $L$.
A differentiable function $\phi$ on a set $X$ is an {\it immersion} if its Jacobian matrix at $x\in X$, denoted by $D\phi(x)$, is injective, $\forall x$.

\subsection{Problem Formulation}
Consider a continuous-time input-affine system given by
\begin{subequations}\label{eq:plant}
\begin{align}
\dot x(t) &= f(x(t)) + g(x(t))u(t),\label{eq:plant_state}\\
y(t) &= h(x(t)) + a(t) + v(t),\label{eq:plant_output}
\end{align}
\end{subequations}
where $x(t)\in\R^n$ is the state,
$u(t)\in\R$ the input,
$y(t)\in\R^p$ the output,
$a(t)\in\R^p$ the attack signal,
and $v(t)\in\R^p$ the measurement noise.
We assume that
$x(t)$, $u(t)$, and $v(t)$ are bounded.
Especially for the state $x(t)$,
let $\XXXX\subset \R^n$ be a compact set,
which is known,
such that $x(t) \in\XXXX$
for all $t\ge 0$.

Regarding the attacker
who generates the signal $a(t)$ and injects
it at the sensors,
we specify the adversarial model next.

{\bf Attack model:} The number of compromised sensors is limited;
there exists a natural number $q<p$ such that up to $q$ components of the attack $a(t)=\col\{a_i(t)\}_{i=1}^{p}$
(out of its $p$ components)
can be non-zero.
Specifically,
the cardinality of the set
\begin{equation}\label{eq:unattacked}
 I_0:=\{i\in [p]: a_i(t)=0, ~\forall t\ge 0\}
\end{equation}
is greater than or equal to $p-q$.
There is no assumptions for the
non-zero components of the attack $a(t)=\{a_i(t)\}_{i=1}^{p}$.
For $i\in [p]\setminus  I_0$,
the attack signal $a_i(t)$ can be arbitrarily designed by an omniscient adversary,
in which,
for example,
the knowledge of all parameters and signals of the system \eqref{eq:plant} can be utilized for the design.
The signal can even be unbounded.

The resilient state estimation problem is stated next.

\begin{prob}\label{prob:estimation}
From the input $u(t)$ and the output $y(t)$ of the system \eqref{eq:plant},
compute a state estimate $\hat x(t)\in\R^n$ such that
\begin{equation}\label{eq:property}
\|\hat x(t) - x(t) \|\le \delta_{\hat x}(t),
\end{equation}
where the estimator error bound $\delta_{\hat x}(t)\ge 0$ is given and does not depend on
the attack signal $a(t)$.
\end{prob}

To achieve the property \eqref{eq:property},
the corrupted components of $y(t)$
need to be identified and excluded from the estimation process, even
though the index set $ I_0$ of the un-corrupted measurements is unknown.

\subsection{Preliminaries}
In this section,
we follow and introduce the approach made in \cite{Kim19TAC},
which designs a high gain observer for each individual channel $y_i(t)$ of the output $y(t) = \col\{y_i(t)\}_{i=1}^{p}$.
Compared with the methods in \cite{Pasqualetti13TAC} or \cite{Chong15ACC},
in which no less than $\binom{p}{q}=\frac{p!}{q!(p-q)!}$ observers are designed against $q$ corrupted measurements out of $p$ measurements,
we construct
comparatively less number of
observers,
as many as the number of measurements.

For the sake of the design of nonlinear observers,
we first make the following assumption.
Although
the system \eqref{eq:plant} is generally not observable from each individual channel $y_i(t)$ of the output $y(t)=\{y_i(t)\}_{i=1}^{p}$,
we assume that a subsystem of \eqref{eq:plant} is uniformly observable,
so that a portion of the state $x(t)$ can be reconstructed from each individual output $y_i(t)$.
\begin{asm}\label{asm:uniform}
For each $i\in[p]$,
the system \eqref{eq:plant_state}
with
the individual output $y_i(t)$
is diffeomorphic to the form
\begin{subequations}\label{eq:decomposition}
		\begin{align}
\begin{split}
		\dot z_i(t) &=
		\col\{\dot z_{i,j}(t)\}_{j=1}^{n_i}\\
		&=
		 \begin{bmatrix} z_{i,2}(t) \\ \vdots \\ z_{i,n_i}(t) \\ \alpha_{i}(z_i(t)) \end{bmatrix}
		+ \begin{bmatrix} \beta_{i,1}(z_{i,1}(t)) \\ \beta_{i,2}(z_{i,1}(t), z_{i,2}(t)) \\ \vdots \\\beta_{i,n_i}(z_{i,1}(t), \dots, z_{i,n_i}(t)) \end{bmatrix}u(t), \end{split}\label{eq:decomposition_obs}
		\\
		\dot z_i'(t) &= F_i'(z_i(t),z_i'(t)) + G_i'(z_i(t),z_i'(t))u(t), \\
		y_i(t) &= z_{i,1}(t) + a_i(t) + v_i(t), \label{eq:decomposition_output}
		\end{align}
	\end{subequations}
	where $z_i(t)\in\R^{n_i}$
	and $z_i'(t)\in\R^{n-n_i}$, with some $n_i\in\N$.
\end{asm}

We note that
the function that maps the state $x(t)\in\R^n$ to the sub-state $z_i(t)\in\R^{n_i}$
is given by the $i$-th component $h_i(\cdot)$
of the output function $h(\cdot)=\col\{h_i(\cdot)\}_{i=1}^{p}$ of \eqref{eq:plant_output},
and its Lie-derivatives along the vector field $f$, as
\begin{equation}\label{eq:phi}
z_i(t) = \begin{bmatrix}
z_{i,1}(t)\\z_{i,2}(t)\\\vdots\\z_{i,n_i}(t)
\end{bmatrix} = \begin{bmatrix}
h_i(x(t))\\ L_fh_i(x(t))\\\vdots\\L_f^{n_i-1}h_i(x(t))
\end{bmatrix}=:\Phi_i(x(t)),
\end{equation}
which can be verified by comparing \eqref{eq:plant} and \eqref{eq:decomposition} with $u(t)\equiv 0$.

Then,
the state $x(t)$ is generally not able to be reconstructed from an individual output $y_i(t)$,
but
the sub-state $z_i(t)=\Phi_i(x(t))$
can be recovered as a piece of $x(t)$,
unless the output $y_i(t)$ is corrupted.
Indeed,
for each $i\in[p]$,
let a high gain observer for the system \eqref{eq:decomposition_obs} be designed, as
\begin{align}
\begin{split}
\dot{\hat{z}}_{i}(t)
&=
\col\{{\dot{\hat z}}_{i,j}(t)\}_{j=1}^{n_i}
\\
&= \begin{bmatrix} {\hat{z}_{i,2}(t)}\\ \vdots \\{\hat{z}_{i, n_i}(t)}\\ \alpha_{i}({\hat{z}_{i}(t)})\end{bmatrix}
+ \begin{bmatrix} \beta_{i,1}({\hat{z}_{i,1}(t)})\\ \beta_{i,2}({\hat{z}_{i,1}(t)}, {\hat{z}_{i,2}(t)}) \\ \vdots \\\beta_{i,n_i}({\hat{z}_{i,1}(t)}, \dots, {\hat{z}_{i, n_i}(t)})\end{bmatrix}u(t)\\
&\qquad -P_{i}^{-1}C_{i}^{T} ( \hat{z}_{i,1}(t)-y_i(t)),\label{eq:observer}
\end{split}
\end{align}
where $\hat z_i(t)\in\R^{n_i}$ is the estimate,
$C_i:=[1,0,\dots,0]\in\R^{n_i}$, and $P_i\in\R^{n_i\times n_i}$
is the unique positive definite solution of
$$ 0 = -\theta_i P_i - A_i^\top P_i - P_i A_i + C_i^\top C_i, $$
in which the parameter $\theta_i$ is to be determined, and
\begin{align*}
A_i&:=\begin{bmatrix}0 & 1 &  \cdots& 0	\\
\vdots &\vdots  &\ddots & \vdots \\
0 & 0 & \cdots& 1	\\
0 & 0 & \cdots& 0	\\	\end{bmatrix}\in \R^{n_i\times n_i}.
\end{align*}
Since $x(t)\in\XXXX$, and the sub-state $z_i(t)=\Phi_i(x(t))$ belongs to the image set $\Phi_i(\XXXX)$ of the function $\Phi_i$,
we let the initial state $\hat z_i(0)$ of \eqref{eq:observer}
be chosen
such that $\hat z_i(0)\in\Phi_i(\XXXX)$.

Now,
the following proposition states that
the observer \eqref{eq:observer} with the $i$-th individual output $y_i(t)$ recovers the information of $z_i(t)=\Phi_i(x(t))$,
in case the output $y_i(t)$ is not corrupted.

\begin{prop}[\!\!\cite{Gauthier92TAC,Kim19TAC}]\label{prop:observer}
There exist $\theta_i^*\ge 1$
and functions
$\eta_i(\theta_i)$ and $\epsilon_i(\theta_i)$ such that, for every $\theta_i\ge \theta_i^*$, the observer \eqref{eq:observer} ensures
\begin{equation}\label{eq:convergence}
\| \hat z_i(t) - z_i(t) \| \le \max\{ 2\eta_i(\theta_i)\|\Phi_i(\XXXX) \|\cdot  e^{- \frac{\theta_i}{8}t} ,\epsilon_i(\theta_i)\}\!=:\!\delta_i(t), 
\end{equation}
provided that $a_i(t)=0$ for all $t\ge 0$.
\end{prop}

{\it Proof:} The proof can be found in \cite{Kim19TAC}.\hfill$\blacksquare$

\begin{rem}
The benefit of constructing an observer for each individual output is that
the majority of the estimates can be preserved,
as many as the number of un-corrupted measurements.
The estimates $\{\hat z_i(t)\}_{i\in I_0}$ from the un-corrupted measurements $\{y_i(t)\}_{i\in I_0}$,
where the set $ I_0$ is such that $| I_0 |\ge p-q$ as defined in \eqref{eq:unattacked},
ensures that \eqref{eq:convergence} holds.
\end{rem}

In the remaining  part of this section,
we formulate a nonlinear equation whose solution will correspond to an estimate for the state $x(t)$.
For each $i\in[p]$,
let us define $e_i(t)$ as the estimation error due to sensor attack so that it is non-zero for the corrupted measurements such that $i\in[p]\setminus I_0$,
and define $r_i(t)$ as the observer error defined from \eqref{eq:convergence} so that it is non-zero for the un-corrupted measurements such that $i\in I_0$.
For $i\in I_0$,
we
define $e_i(t):=0\in\R^{n_i}$
and
$r_i(t):=\hat z_i(t) - z_i(t)\in\R^{n_i}$, so that
the following equality
\begin{equation}\label{eq:static_individual}
\hat z_i(t) = \Phi_i(x(t))+r_i(t) +  e_i(t)
\end{equation}
holds for $i\in I_0$, where
the inequality \eqref{eq:convergence} implies $\|r_i(t)\|\le \delta_i(t)$.
And,
for $i\in[p]\setminus I_0$,
we define $r_i(t):=0\in\R^{n_i}$
and
$e_i(t):=\hat z_i(t) - z_i(t)\in\R^{n_i}$ so that \eqref{eq:static_individual} holds for $i\in[p]\setminus I_0$.

Now,
we rewrite the equations \eqref{eq:static_individual} for all $i\in[p]$ at once, as
\begin{equation}\label{eq:static}
\hat z(t) = \Phi(x(t)) + r(t) + e(t),
\end{equation}
in which
\begin{align*}
\begin{split}
\hat z(t)&:= \col\{\hat z_i(t)\}_{i=1}^{p},\\
\Phi(x(t))&:=\col\{\Phi_i(x(t))\}_{i=1}^{p},
\end{split}\quad
\begin{split}
r(t)&:=\col\{r_i(t)\}_{i=1}^{p},\\
e(t)&:=\col\{e_i(t)\}_{i=1}^{p}
\end{split}
\end{align*}
are defined as the stacks of partitioned vectors
as elements of the space $\R^\NN$ of partitioned vectors where $\NN = (n_1,\dots,n_p)$.
Here,
the vector $r(t)\in\R^\NN$ is defined such that $\| r(t)\|\le
\delta(t):=
\max_{i\in [p]}\{\delta_i(t)\}$, and the vector $e(t)\in\R^\NN$ is defined such that $\|e(t)\|_0^\NN\le q$, i.e.,
not more than $q$ elements of $e(t)=\col\{e_i(t)\}_{i=1}^{p}$ can be non-zero, out of the $p$ elements.

Then,
it can be understood that
the resilient state estimation problem is to solve the equation \eqref{eq:static}
and find the value of $x(t)$,
from the knowledge of the estimates $\hat z(t)\in\R^\NN$ together with
the functions $\Phi(\cdot)=\col\{\Phi_i(\cdot)\}_{i=1}^{p}$
and the constraints
\begin{equation}\label{eq:constraint}
\|r(t)\|\le \delta(t),\qquad \|e(t)\|_0^\NN\le q.
\end{equation}

In \eqref{eq:static}, note that
$e(t)\in\R^\NN$ denotes the error due to the attack injection,
and
$r(t)\in\R^\NN$ denotes the transient errors of the observers plus the effect of the noise corruption.
Indeed, if there is no compromised sensors so that $ I_0=[p]$, then $e(t)=0$, by construction.
And, if the effect of the noise $v_i(t)$ in \eqref{eq:decomposition_output} and
the transient errors of the observers are ideally zero, so that the error bound $\delta_i(t)$ in \eqref{eq:convergence} is set as zero and the un-compromised estimates $\{\hat z_i(t)\}_{i\in I_0}$ are equal to the sub-states $\{z_i(t)\}_{i\in I_0}$,
then we have $r(t)=0$.

\section{Main Results}

We now solve \eqref{eq:static}
and find an approximate value of $x(t)$
for the resilient state estimation problem.
First, the problem of {\it attack identification} is stated as follows.

\begin{prob}\label{prob:identification}
Given the information
$\{\hat z_i(t)\}_{i=1}^{p}$ and
$\{\Phi_i\}_{i=1}^{p}$ and \eqref{eq:static_individual},
find an index set $I\subset[p]$, $|I|=p-\!q$, such that
$ \|\hat z_i(t) - \Phi_i(x(t))\| \le \delta_{\hat z}(t)$ holds,
$\forall i\in I$,
where
the
bound $\delta_{\hat z}(t)\ge 0$
does not depend on
the errors
$\{e_i(t)\}_{i=1}^{p}$.
\end{prob}

A solution to the attack identification can yield a solution to the resilient state estimation;
if a set $\{\hat z_i(t)\}_{i\in I}$, $I\subset [p]$, is a solution to Problem~\ref{prob:identification}
and if there is a Lipschitz function $\psi$ such that $\psi(\col\{\Phi_i(x)\}_{i\in I})=x$, $\forall x\in\XXXX$, then
it is obvious that
$\hat x(t)= \psi(\col\{\hat z_i(t)\}_{i\in I})$ becomes a solution to Problem~\ref{prob:estimation}.

Now, we note that the number of cases considered for attack identification 
with all the estimates $\{\hat z_i(t)\}_{i=1}^p$,
or resilient state estimation
with respect to all the measurements,
is $\binom{p}{q}$. Hence the problem has combinatorial complexity. For example, see \cite{Kim19TAC,Lee20TAC,Chong15ACC,Fawzi14}, or Section~\ref{subsec:redundancy}.

\begin{figure*}
	\centering
\includegraphics[width=0.9\textwidth]{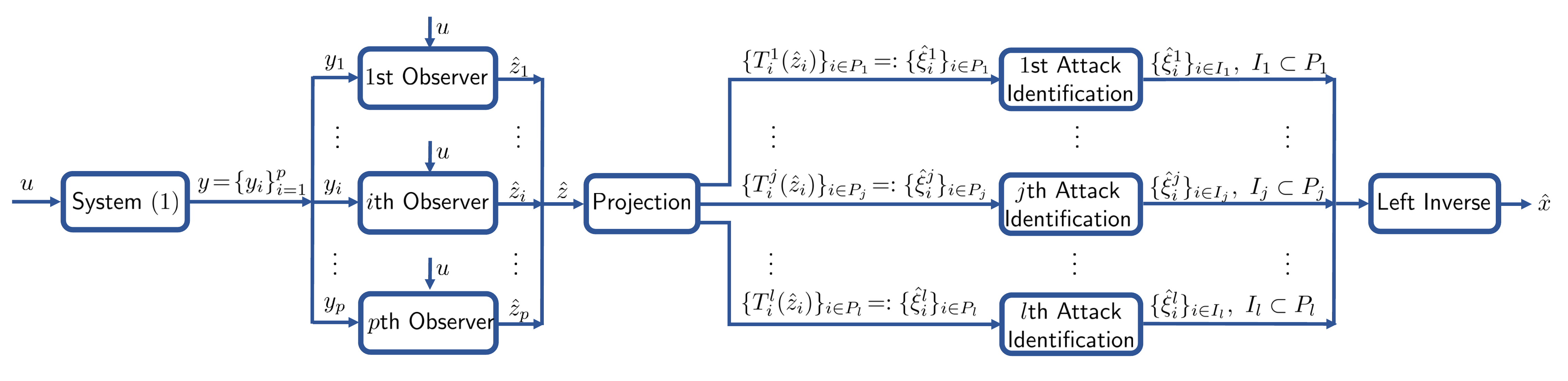}
	\caption{Configuration of the proposed scheme as three steps: i) partial state observer design, ii) attack identification with projected local estimates, and iii) state reconstruction with identified estimates.
	}
	\label{fig:config}
\end{figure*}

To reduce the complexity, we propose {\it the attack identification is performed locally}.
As described in Fig.~\ref{fig:config},
the estimate set $\{\hat z_i(t)\}_{i=1}^p$ is proposed to be decomposed into local estimates
$$ \{T_i^j(\hat z_i(t))\}_{i\in P_j}=:\{\hat \xi_i^j(t)\}_{i\in P_j},\quad P_j\subset [p],~~j=1,\dots,l,$$
with some projection $T_i^j$
for each $i\in P_j$.
Consider Problem~\ref{prob:identification} with
the
sets $\{\hat \xi_i^j(t)\}_{i\in P_j}$ and $\{T_i^j\circ \Phi_i\}_{i\in P_j}$,
with the equation
$$\hat \xi_i^j(t) = T_i^j\circ \Phi_i (x(t)) + r_i^j(t) + e_i^j(t),$$
where $e_i^j(t) = 0$ and $r_i^j = \hat \xi_i^j(t) - T_i^j(\Phi_i (x(t)))$ for $i\in I_0$,
and $r_i^j(t) = 0$ and $e_i^j = \hat \xi_i^j(t) - T_i^j(\Phi_i (x(t)))$ for $i\in P_j\setminus I_0$.
If there exists a solution to the attack identification with respect to each $j$-th local set 
and if the state $x(t)$ can be reconstructed from the identified local estimates, then
the the number of cases will become $\sum_{j=1}^l \binom{|P_j |}{q}$,
which is significantly smaller than $\binom{p}{q}$ if the set $P_j$ for each $j$ is a proper subset of $[p]$.

To this end,
in Section~\ref{subsec:redundancy},
we introduce a condition for redundancy, under which the set of un-compromised estimates can be identified,
regarding a given set of estimates $\{\hat z_i(t)\}$ with corresponding functions $\{\Phi_i(\cdot)\}$.
In Section~\ref{subsec:estimation},
the method for resilient state estimation,
i.e.,
for solving \eqref{eq:static}, is proposed, based on local attack identification.
The introduced redundancy condition will be compared to the existing redundant observability condition,
and the complexity of the problem will be discussed.
Then,
in Section~\ref{subsec:cases},
we introduce a class of systems which includes all linear systems, where the proposed method
can be implemented in a particularly constructive way.
The case of linear systems will be illustrated as an example.

\subsection{Redundancy Notion for Attack Identification}\label{subsec:redundancy}

We first introduce the notion of redundancy.
Consider a set $\Phi(x)=\{\Phi_i(x)\}_{i=1}^{p}\in\R^\NN$ of partial information of the state $x\in\XXXX$,
and
let a subset 
$\{\Phi_i(x)\}_{i\in I}$ of
$\{\Phi_i(x)\}_{i=1}^{p}$
be chosen,
with respect to an index subset $ I\subset[p]$.
Then,
the rest portion $\{\Phi_i(x)\}_{i\in[p]\setminus I}$
can be regarded as
{\it redundant},
if the information of $\{\Phi_i(x)\}_{i=1}^{p}$ can be restored from the portion $\{\Phi_i(x)\}_{i\in I}$.
The definition is formally stated, as follows.

\begin{defn}\label{def:redundant}
A map
$\Phi:\XXXX\ra\R^\NN$ with $|\NN|=p$
is called {\it $k$-redundant},
if for any $ I\subset[p]$ such that $| I|\ge p-k$, the map
$$\pi_I:~\Phi(x)=\col\{\Phi_i(x)\}_{i=1}^{p}~\mapsto~\Phi_I(x)= \col\{\Phi_i(x)\}_{i\in I}$$
is injective on
the image set
$\Phi(\XXXX)$;
i.e.,
there exists a function $\pi_I^{-1}$
such that
$\pi_I^{-1}(\Phi_I(x))
=\Phi(x)$ holds, for all $x\in\XXXX$.
\end{defn}

\begin{figure}[!h]
	\centering
\includegraphics[width=0.2\textwidth]{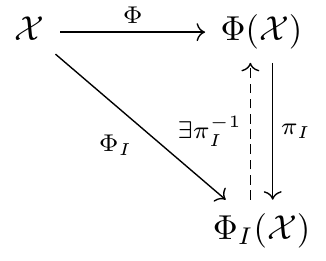}\\
\includegraphics[width=0.35\textwidth]{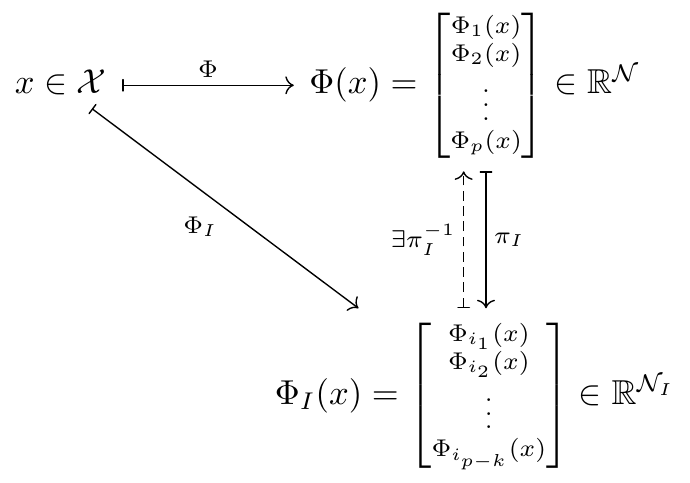}
	\caption{Commutative diagram for a $k$-redundant map, $k\in\N$.
	}
	\label{fig:commutative}
\end{figure}

Since injectivity
is equivalent to
left-invertibility,
note that the injectivity of the projection $\pi_I$ restricted to the image space $\Phi(\XXXX)$ implies the existence of the left-inverse\footnote{
Rigorously,
the left inverse of the restriction of $\pi_I$ to the set $\Phi(\XXXX)$ should be denoted as $(\pi_I|\Phi(\XXXX))^{-1}$,
as the
the projection $\pi_I$ defined on $\R^\NN$ itself is not left-invertible.
Nonetheless, we abuse notation and use $\pi_I^{-1}$, for simplicity.
} $\pi_I^{-1}$,
satisfying
$$\pi_I^{-1}(\pi_I(\Phi(x)))=\pi_I^{-1}(\Phi_I(x))=\Phi(x)$$ for all $x\in\XXXX$.
Thus,
the $k$-redundancy
of $\Phi(x)=\{\Phi_i(x)\}_{i=1}^{p}$
means that the information
can be restored, even if $k$ elements of $\Phi(x)$ are removed by
the projection $\pi_I$.
Fig.~\ref{fig:commutative} shows a commutative diagram describing the definition of $k$-redundancy.

Then,
a condition equivalent to the redundancy is provided as follows,
which will be used for the attack identification.

\begin{prop1}\label{prop:characterization_ideal}
For a function $\Phi:\XXXX\ra\R^\NN$,
the following statements are equivalent:
\begin{enumerate}
\item
For $x_1\in\XXXX$, $x_2\in\XXXX$, and $e\in\R^\NN$ such that $\|e\|_0^\NN\le k$,
$\Phi(x_1)=\Phi(x_2)+e$ implies
$\Phi(x_1)=\Phi(x_2)$ and
$e=0$.
\item The function $\Phi$ is $k$-redundant.\hfill$\square$
\end{enumerate}
\end{prop1}

{\it Proof:} 1) $\Rightarrow$ 2):
Let $|\NN|=p$, and
let $ I\subset[p]$ be an index set such that $| I|\ge p-k$.
We show that
the projection
$\pi_I:\Phi(x)
\mapsto\Phi_I(x)$ is injective on  $\Phi(\XXXX)$.
Suppose $\Phi_I(x_1)=\Phi_I(x_2)$ for some $x_1\in\XXXX$ and $x_2\in\XXXX$. Let $e=\Phi(x_1)-\Phi(x_2)$.
It is obvious that $\|e\|_0^\NN\le k$ and $\Phi(x_1)=\Phi(x_2)+e$,
and hence $\Phi(x_1)=\Phi(x_2)$.
Thus, $\pi_I$ is injective on $\Phi(\XXXX)$, and $\Phi$ is $k$-redundant.
2) $\Rightarrow$ 1): Let $\Phi(x_1)=\Phi(x_2)+e$.
Since $\|e\|_0^\NN\le k$, there exists $ I\subset[p]$
such that $| I|\ge p-k$
where $p=|\NN|$, which satisfies $\Phi_I(x_1)=\Phi_I(x_2)$.
By Definition~\ref{def:redundant},
it follows that $\Phi(x_1)=\Phi(x_2)$.
The proof is completed.
\hfill$\blacksquare$

The condition 1) in Proposition~\ref{prop:characterization_ideal} means that
any set of data
$\{\Phi_i(x)\}_{i=1}^{p}+e$ corrupted by $e\in\R^\NN$ cannot be an element of the image space $\Phi(\XXXX)$ other than $\Phi(x)=\{\Phi_i(x)\}_{i=1}^{p}$,
as long as no more than $k$ elements are compromised, i.e., $\|e\|_0^\NN\le k$.
Then,
given the vector $\Phi(x)+e$,
it follows that
the presence of attack $e$ can be inspected by checking the condition $\Phi(x)+e\in\Phi(\XXXX)$, since 
$e=0$
if and only if $\Phi(x)+e\in\Phi(\XXXX)$.

Then,
given the equation \eqref{eq:static} with \eqref{eq:constraint},
we first consider the case $\delta(t)=0$,
i.e., the ideal case that there is no effect of the noise and the transient observer errors in \eqref{eq:convergence},
and
suggest our observation applied to identify a subset of un-compromised estimates,
among the estimates $\hat z(t)=\{\hat z_i(t)\}_{i=1}^{p}$.
Indeed,
for an index set $ I\subset[p]$ with $| I|=p-q$,
let
\eqref{eq:static} be reduced as
\begin{equation} \label{eq:static_reduced}
\hat z_I(t)= \Phi_I(x(t)) + e_I(t),
\end{equation}
where $\hat z_I(t):=\col\{\hat z_i(t)\}_{i\in I}$ and $e_I(t):=\col\{e_i(t)\}_{i\in I}$.
Then,
the estimates $\hat z_I(t)$ can be regarded as un-compromised 
if $e_I(t)=0$,
and it can be inspected by checking
whether the vector $\hat z_I(t)$ belongs to the image of $\Phi_I$, i.e., $\hat z_I(t)\in\Phi_I(\XXXX)$.
Since there is up-to $q$ compromised estimates
and there are $\binom{p}{q}$ cases for choosing the subset $ I\subset[p]$,
the case $e_I(t)=0$ can be found by inspecting the $\binom{p}{q}$ cases one by one,
and
it will be able to restore 
the information of $\Phi(x(t))$,
under redundancy.
This observation is stated as the following proposition.
\begin{prop1}\label{prop:identification_ideal}
Consider \eqref{eq:static} subject to \eqref{eq:constraint}. Assume that $\delta(t)=0$, and that the function
$\Phi$ is $2q$-redundant.
\begin{enumerate}
\item Let
$ I\subset[p]$ be an index subset
such that $| I|=p-q$. Then
$e_I(t)=0$ if and only if $\hat z_I(t)\in\Phi_I(\XXXX)$.
\item There exists
$ I\subset[p]$
with
$| I|=p-q$ such that it satisfies $\hat z_I(t)\in\Phi_I(\XXXX)$
and
$\pi_I^{-1}(\hat z_I(t))=\Phi(x(t))$.
\end{enumerate}
\end{prop1}

{\it Proof:} 1)
The function
$\Phi_I$ with $ I\subset[p]$, $| I|=p-q$, is $q$-redundant,
since $\Phi$ is $2q$-redundant.
Then, given $\hat z_I(t)=\Phi_I(x(t))+e_I(t)$,
by Proposition~\ref{prop:characterization_ideal},
$e_I(t)=0$
if and only if
there exists $x'\in\XXXX$ such that $\Phi_I(x')=\hat z_I(t)$,
i.e.,
$\hat z_I(t)\in\Phi_I(\XXXX)$.
2) Since $\|e(t)\|_0^\NN\le q$, there exists $ I\subset[p]$, $| I|=p-q$, such that $e_I(t)=0$ and $\hat z_I(t)=\Phi_I(x(t))$.
It follows that
$\pi_I^{-1}(\hat z_I(t))=\Phi(x(t))$,
because $\Phi$ is $q$-redundant.
 \hfill$\blacksquare$

Now,
for the general case $\delta(t)\neq0$ in \eqref{eq:static} and \eqref{eq:constraint},
to take into account the presence of the observer errors in \eqref{eq:convergence} as well,
we introduce
a stronger version of Definition~\ref{def:redundant}
as follows,
which also requires
the Lipschitz property\footnote{In fact, thanks to  compactness of the domain such as $\XXXX$ or $\Phi(\XXXX)$, the Lipschitz property can be ensured from the smoothness of the functions.} for each left-inverse $\pi_I^{-1}:\Phi_I(x)\mapsto\Phi(x)$ of the projection $\pi_I$ restricted to $\Phi(\XXXX)$.

\begin{defn}\label{def:Lipschitz_redundant}
A $k$-redundant map $\Phi:\XXXX\ra\R^\NN$ is called
{\it $k$-Lipschitz redundant},
if there exists a constant $M_\Phi$ such that $$\|\Phi(x_1)-\Phi(x_2) \|\le M_\Phi\|\Phi_I(x_1)-\Phi_I(x_2)\|,\quad \forall x_1,x_2\in\XXXX,$$
for all
$ I\subset[p]$ such that $| I|\ge p-k$, where $p=|\NN|$.
\end{defn}

The constant $M_\Phi$ can be regarded as a Lipschitz constant of
the map
$\pi_I^{-1}:\Phi_I(x)\mapsto \Phi(x)$, for all $I\subset[p]$ with $|I|\ge p-k$.

Analogous to Proposition~\ref{prop:characterization_ideal},
an equivalent condition of the Lipschitz redundancy is stated, as follows.

\begin{lem1}\label{lem:characterization}
For a function $\Phi:\XXXX\ra\R^\NN$, the followings are equivalent:
\begin{enumerate}
\item There exists
a constant
$M_\Phi\ge 0$
such that 
$$\|\Phi(x_1)-\Phi(x_2)\|\le M_\Phi\|\Phi(x_1)-\Phi(x_2)+e\|,$$
for all $x_1\!\in\!\XXXX$, $x_2\!\in\!\XXXX$, and $e\!\in\!\R^\NN$ such that $\|e\|_0^\NN\le k$.

\item The function $\Phi$ is $k$-Lipschitz redundant.\hfill$\square$
\end{enumerate}
\end{lem1}

{\it Proof:} 1) $\Rightarrow$ 2):
Let $p=|\NN|$.
For $\Phi(x_1)=\col\{\Phi_i(x_1)\}_{i=1}^{p}$, $\Phi(x_2)=\col\{\Phi_i(x_2)\}_{i=1}^{p}$,
and $ I\subset[p]$ with $| I|\ge p-k$,
let $e=\col\{e_i\}_{i=1}^{p}$ be such that
$e_i=\Phi_i(x_2)-\Phi_i(x_1)$ for $i\in[p]\setminus I$,
and
$e_i=0$ for $i\in I$.
Then, $\|e\|_0^\NN\le k$,
and $\|\Phi(x_1)-\Phi(x_2)+e\|=\|\Phi_I(x_1)-\Phi_I(x_2)\|$.
Hence,
$\|\Phi(x_1)-\Phi(x_2) \|\le M_\Phi\|\Phi_I(x_1)-\Phi_I(x_2)\| $ holds.
2) $\Rightarrow$ 1):
Given
$x_1\in\XXXX$,
$x_2\in\XXXX$,
and $e=\col\{e_i\}_{i=1}^{p}$ with $\|e\|_0^\NN\le k$,
let
$ I\subset[p]$, $| I|\le p-k$, be such that
$e_I=0$,
i.e.,
$e_i=0$, $\forall i\in I$.
Note that
$\|z_I\|\le \|z\|$,
for all $z=\col\{z_i\}_{i=1}^{p}$ and
$z_I=\col\{z_i\}_{i\in I}$.
Then, putting $z=\Phi(x_1)-\Phi(x_2)+e$,
it follows that
$
\|\Phi(x_1)-\Phi(x_2)\|\le M_\Phi\|\Phi_I(x_1)-\Phi_I(x_2)\|
\le M_\Phi\|\Phi(x_1)-\Phi(x_2)+e \|
$
with some $M_\Phi\ge 0$,
since $\Phi$ is $k$-Lipschitz redundant.
\hfill$\blacksquare$

Now, our method for the attack identification is proposed,
which takes account of the general case of \eqref{eq:static} with the observer errors $\delta(t)\neq0$.
Recall that, given a vector $\hat z_I(t)$ that obeys \eqref{eq:static_reduced}, where $\delta(t)=0$,
the method of Proposition~\ref{prop:identification_ideal}
checks
if
$\hat z_I(t)$ belongs to the set $\Phi_I(\XXXX)$,
to inspect the presence of the attack.
Then,
we propose that
the method can be extended,
by measuring the distance
$d(\hat z_I(t),\Phi_I(\XXXX))$
from $\hat z_I(t)$ to $\Phi_I(\XXXX)$
and checking the condition
\begin{equation}\label{eq:inspection}
d(\hat z_I(t),\Phi_I(\XXXX))\le \delta(t),
\end{equation}
when $\Phi$ is Lipschitz-redundant,
as the following theorem.

\begin{thm1}\label{thm:identification}
Consider \eqref{eq:static} subject to \eqref{eq:constraint}, and assume that the function $\Phi$ is $2q$-Lipschitz redundant.
\begin{enumerate}
\item\label{item:thm1.1}
Let $ I\subset[p]$ with $| I|=p-q$ be an index set.
If \eqref{eq:inspection} is violated, then $e_I(t)\neq0$.
If \eqref{eq:inspection} holds,
then
\begin{equation}\label{eq:attack_free}
\|\hat z_I(t)-\Phi_I(x(t))\|\le (2M_\Phi+1)\delta(t).
\end{equation}
\item\label{item:thm1.2} There exists
$ I\subset[p]$
with
$| I|=p-q$,
such that
\eqref{eq:inspection} and
\begin{equation}\label{eq:recover}
\|\psi_I(\hat z_I(t)) - \Phi(x(t))  \|\le 
(2M_\Phi^2+M_\Phi)\delta(t)
\end{equation}
holds,
with
a Lipschitz map $\psi_I$, with
$L(\psi_I)= M_\Phi$.
\hfill$\square$
\end{enumerate}
\end{thm1}

Before the proof,
the meaning of Theorem~\ref{thm:identification}
is understood as follows.
If \eqref{eq:inspection} is violated with a set $\hat z_I(t)=\{\hat z_i(t)\}_{i\in I}$ of estimates,
then $e_I(t)\neq0$;
i.e.,
there is a compromised element in the set $\hat z_I(t)$ being inspected.
Otherwise,
if \eqref{eq:inspection} holds,
then
\begin{align*}
\| \hat z_I(t) - \Phi_I(x(t))\|=\|e_I(t)+r_I(t)\|\le (2M_\Phi+1)\delta(t),
\end{align*}
which implies that
the error $\| \hat z_I(t) - \Phi_I(x(t))\|$
and
the effect  $e_I(t)$ of the attack is small,
and cannot be distinguished
from the error $r_I(t)$ due to noises.
And, Theorem~\ref{thm:identification}.\ref{item:thm1.2} means that
a set $\hat z_I(t)$ satisfying \eqref{eq:inspection} will be found by inspecting the cases one by one,
and the information of $\Phi(x)$ can be recovered from $\hat z_I(t)$ with the function $\psi_I$,
which will be seen as a Lipschitz extension of the function
$\pi_I^{-1}$ defined on the set $\Phi_I(\XXXX)$.

{\it Proof of Theorem\,\ref{thm:identification}:}
1):
Clearly,
$e_I(t)=0$ implies $\hat z_I(t)=\Phi_I(x(t))+r_I(t)$,
which
is followed by
$\|\hat z_I(t)-\Phi_I(x(t)) \|\le \delta(t)$, i.e., \eqref{eq:inspection} holds.
Now, suppose that \eqref{eq:inspection} holds.
As $\XXXX$ is compact,
there exists $x'\in\XXXX$
such that $\|\hat z_I(t) - \Phi_I(x')\| \le \delta(t)$.
Then,
since
$\Phi$ is $2q$-Lipschitz redundant,
$\Phi_I$ is $q$-Lipschitz redundant,
the equation \eqref{eq:static} implies
$\hat z_I(t) = \Phi_I(x(t))+ r_I(t)+ e_I(t)$,
and
$\|e_I(t)\|_0^\NN\le q$,
it follows
that
\begin{align}
\begin{split}
&\| \hat z_I(t) - \Phi_I(x(t))\|\\
&\quad\le \|\hat z_I(t) - \Phi_I(x')\| + \|\Phi_I(x(t)) - \Phi_I(x') \|\\
&\quad\le \delta(t) + M_\Phi\|\Phi_I(x(t))-\Phi_I(x')+e_I(t)\|\\
&\quad\le \delta(t) + M_\Phi(\|\hat z_I(t) - \Phi_I(x') \| + \| r_I(t)\|)\\
&\quad\le (2M_\Phi+1)\delta(t),\label{eq:Thm1}
\end{split}
\end{align}
by Lemma~\ref{lem:characterization}.
Thus, 1) is proved.
2): Since $\|e(t)\|_0^\NN\le q$, there exists $ I\subset[p]$ with $| I|\le p-q$ such that $e_I(t)=0$ and $\hat z_I(t)=\Phi_I(x(t))+r_I(t)$. It follows that \eqref{eq:inspection} and \eqref{eq:attack_free} hold.
Now, since $\Phi_I$ is $q$-Lipschitz redundant,
by Definitions~\ref{def:redundant} and
\ref{def:Lipschitz_redundant},
it can be checked that
the map $\pi_I^{-1}:\Phi_I(x)\mapsto\Phi(x)$ defined on the set $\Phi_I(\XXXX)
$ is Lipschitz with the constant $M_\Phi=L(\pi_I^{-1})$.
Then, according to Kirszbraun's Lipschitz extension theorem\footnote{For a Lipschitz function
$\phi:X \ra \R^n$,
a Lipschitz extension of each component $\phi_i(\cdot)$ of $\phi(\cdot)=\col\{\phi_i(\cdot)\}_{i=1}^{n}$
can be found as
$$\overline \phi_i(x):= \inf_{x'\in X}\{ \phi_i(x')+ L(\phi)\|x-x' \| \}.$$
See \cite[p.~21]{Schwartz69} for more details.
},
there exists a Lipschitz function $\psi_I$ that satisfies
$\psi_I(\Phi_I(x))=\pi_I^{-1}(\Phi_I(x))$
for every 
$x\in\XXXX$,
and
satisfies
$\|\psi_I( z_I)-\psi_I( z'_I)\|\le M_\Phi\|  z_I -  z'_I \| $ for any real vectors $ z_I$ and $ z'_I$.
Then, by putting $ z_I =\hat z_I(t)$ and $z_I' = \Phi_I(x(t))$,
the proof is completed, by the inequality \eqref{eq:attack_free}.
\hfill$\blacksquare$

We have proposed that the inspection and identification of the attacks can be performed by checking if the inequality \eqref{eq:inspection} holds or not, for each $ I\subset[p]$ such that $| I|=p-q$.
Regarding the implementation in practice, one may wonder how the distance $d(\hat z_I(t),\Phi_I(\XXXX))$ can be computed,
despite that the compactness of the set $\Phi_I(\XXXX)$ guarantees its existence.
One way would be to consider a set of sampled data for the set $\Phi_I(\XXXX)$;
let $\XXXX_\Delta\subset\XXXX$ be a finite subset of $\XXXX$ such that with some constant $\Delta>0$, it satisfies $d(x,\XXXX_\Delta)\le \Delta$ for all $x\in\XXXX$,
i.e.,
for any $x\in\XXXX$, there exists an element $x'\in \XXXX_\Delta$ such that $\| x-x'\| \le \Delta$.
Then,
an approximate value of the distance $d(\hat z_I(t),\Phi_I(\XXXX))$
for checking \eqref{eq:inspection} can be obtained by
\begin{equation*}
d(\hat z_I(t), \Phi_I(\XXXX_\Delta))= \min_{z'\in\Phi(\XXXX_\Delta)} \|\pi_I(z')- \hat z_I(t) \|,
\end{equation*}
which can be computed from the information of the finite set $\Phi_I(\XXXX_\Delta)=\{\Phi_I(x)\}_{x\in\XXXX_\Delta}$.
Then, the following corollary states that
the attack identification can also be performed with the approximate distance measured with the sampled data.

\begin{cor}\label{cor:sample}
Let $ I\subset[p]$ with $| I|=p-q$ be given.
If
\begin{equation}\label{eq:inspection_approx}
d(\hat z_I(t),\Phi_I(\XXXX_\Delta))\le \delta(t)+L(\Phi_I)\Delta
\end{equation}
is violated,
then $e_I(t)\neq 0$.
Otherwise,
if \eqref{eq:inspection_approx} holds, then
$$\|\hat z_I(t) - \Phi_I(x(t)) \|\le (2M_\Phi +1)\delta(t)+(M_\Phi+1)L(\Phi_I)\Delta$$ holds.
\end{cor}

{\it Proof:}
Suppose $e_I(t)=0$ so that $\hat z_I(t)=\Phi_I(x(t))+r_I(t)$.
Since $\| \Phi_I(x(t))+r_I(t) - \Phi_I(x') \|\le  L(\Phi_I)\|x(t) - x' \|+\delta(t)$ for any $x'\in\XXXX_\Delta$,
it follows that \eqref{eq:inspection_approx} holds.
Next, suppose that \eqref{eq:inspection_approx} holds.
Then, as analogous to \eqref{eq:Thm1},
it can be verified that
\begin{align*}
&\| \hat z_I(t) - \Phi_I(x(t))\|\\
&\quad\le \|\hat z_I(t) - \Phi_I(x')\| + \|\Phi_I(x(t)) - \Phi_I(x') \|\\
&\quad\le (\delta(t)+L(\Phi_I)\Delta)+ M_\Phi(2\delta(t)+L(\Phi_I)\Delta)
\end{align*}
with some $x'\in\XXXX_\Delta$. This completes the proof.\hfill$\blacksquare$

Finally,
we provide
a way for checking the ``Lipschitzness'' of a given redundant function,
which may be easier to check, compared with the conditions in Definition~\ref{def:Lipschitz_redundant} or Lemma~\ref{lem:characterization}.

\begin{prop}
Let $\Phi(\cdot)=\col\{\Phi_i(\cdot)\}_{i\in[p]}$ be a $k$-redundant function.
If the rank of the Jacobian matrix $D\Phi(x)$ of $\Phi$ at each $x\in\XXXX$ is the same as that of $D\Phi_I(x)$, for all $ I\subset[p]$ with $| I|\ge p-k$,
then $\Phi$ is $k$-Lipschitz redundant.
\end{prop}

{\it Proof:}
Suppose that $\Phi$ is
not $k$-Lipschitz redundant;
with some $ I\subset[p]$ such that $| I|\ge p-k$,
let $\{x_i\}_{i=1}^{\infty}$ and $\{x'_i\}_{i=1}^{\infty}$ be sequences in $\XXXX$ such that
$\Phi(x_i)\neq \Phi(x_i')$, $\forall i=1,\dots$, and
\begin{equation}\label{eq:prop4}
\lim_{i \ra \infty}\frac{\| \Phi_I(x_i) - \Phi_I(x_i') \|}{\|\Phi(x_i) - \Phi(x_i') \|}=0.
\end{equation}
Since $\XXXX$ is compact,
by Bolzano-Weierstrass theorem, we may assume that $\{x_i\}_{i=1}^{\infty}$ and $\{x_i'\}_{i=1}^{\infty}$ converge
to points
$x_\infty$ and $x_\infty'$ in $\XXXX$,
respectively.
If $\Phi(x_\infty) \neq \Phi(x_\infty')$, then $\Phi_I(x_\infty) \neq \Phi_I(x_\infty')$ as $\Phi$ is $k$-redundant and $\pi_I$ is injective on $\Phi(\XXXX)$,
and it contradicts to \eqref{eq:prop4}.
If $\Phi(x_\infty) = \Phi(x_\infty')$,
then we have
$$\lim_{i \ra \infty}\!\! \frac{\|\Phi_I(x_i)\!-\!\Phi_I(x_i')\!-\!D\pi_I(\Phi(x_\infty))\!\cdot\!(\Phi(x_i)\!-\!\Phi(x_i'))\|}{\|\Phi(x_i)-\Phi(x_i')\|}=0 $$
by
continuous differentiability
of $\pi_I$,
where $D\pi_I(\Phi(x_\infty))$ is the Jacobian matrix of $\pi_I$ at $\Phi(x_\infty)$.
And,
it is followed by
\begin{equation}\label{eq:prop4_}
\lim_{i \ra \infty}
\left\|D\pi_I(\Phi(x_\infty))\cdot \frac{\Phi(x_i)-\Phi(x_i')}{\|\Phi(x_i)-\Phi(x_i')\|}\right\|=0,
\end{equation}
by \eqref{eq:prop4}.
Now, from
$\Phi_I = \pi_I\circ \Phi$,
note that
$D\Phi_I(x_\infty) = D\pi_I(\Phi(x_\infty))\cdot D\Phi(x_\infty)$.
It implies that
the rank of $D\Phi_I(x_\infty)$ is equal to that of $D\Phi(x_\infty)$, so the linear map $D\pi_I(\Phi(x_\infty))$ is injective on the image of $D\Phi(x_\infty)$.
It contradicts to \eqref{eq:prop4_}, and hence,
$\Phi$ is $k$-Lipschitz redundant.
The proof is completed.
\hfill$\blacksquare$

So far,
given the estimates $\{\hat z_i(t)\}$ with the functions $\{\Phi_i\}$,
we have introduced the redundancy notion,
under which
the un-corrupted estimates can be identified and partial information $\{\Phi_i(x(t))\}$ of the state $x(t)$ can be restored without the effect of attack.
In the next sections,
it will be noted that
the identification is possible for local subsets of the estimates in case they
are also redundant,
and we continue to recover the state $x(t)$, to perform the resilient state estimation.
Conditions for the local identification are to be studied, and
benefits of such cases are to be discussed,
regarding computational efforts.

\subsection{Resilient State Estimation}\label{subsec:estimation}
We begin with
a corollary of Theorem~\ref{thm:identification} as follows,
which shows that 
the recovery of the state $x(t)$ is possible,
when
the map $\Phi$ is left-invertible as well as $2q$-Lipschitz redundant.

\begin{cor}\label{cor:centralized}
If there exists a Lipschitz function $\Phi^{-1}$ such that $\Phi^{-1}(\Phi(x))=x$ for all $x\in\XXXX$, then \eqref{eq:recover} implies that
\begin{equation}\label{eq:recover_state}
\|\Phi^{-1}(\psi_I(\hat z_I(t))) - x(t)  \|\le 
L(\Phi^{-1})(2M_\Phi^2+M_\Phi)\delta(t)
\end{equation}
holds, where $L(\Phi^{-1})$ is a Lipschitz constant of $\Phi^{-1}$.
\end{cor}

{\it Proof:}
The proof is straightforward.
\hfill$\blacksquare$

With the identified partial estimate $\hat z_I(t)$ from Theorem~\ref{thm:identification}
and the left inverse of $\Phi$,
the estimate $\Phi^{-1}(\psi_I(\hat z_I(t)))$ for $x(t)$
is clearly a solution to the resilient state estimation,
as
its error bound is determined by the function $\Phi$ and the bound $\delta(t)$ for the noise, which does not depend on the attack $e(t)$.

A couple of remarks are made as follows, regarding the required condition for the method of Corollary~\ref{cor:centralized}.

\begin{rem}\label{rem:immersion}
The Lipschitz left inverse of $\Phi$ exists, if the map $\Phi$ is an injective immersion; i.e.,
it is injective and its Jacobian $D\Phi(x)$ is injective at every $x\in\XXXX$. See \cite[Proposition 1]{Kim19TAC}.
\end{rem}

\begin{rem}\label{rem:red_obs}
The required condition,
the combination of $2q$-Lipschitz redundancy of $\Phi$ and its left-invertibility
is equivalent to the existing ``$2q$-redundant observability,''
which is commonly considered as necessary condition for the resilient state estimation, as investigated in \cite{Pasqualetti13TAC,Fawzi14,Shoukry16TAC,Chong15ACC,Chanhwa19TAC,Kim19TAC}.
Indeed,
the assumption made
in \cite{Kim19TAC} is that
each subset of functions $\Phi_I=\{\Phi_i\}_{i\in I}$ of $\Phi=\{\Phi_i\}_{i\in[p]}$ with respect to $ I\subset[p]$, $| I|=p-2q$,
is injective and has a left inverse which is Lipschitz, and this is in fact equivalent\footnote{
As long as it is assumed that
the system \eqref{eq:plant} is observable and
$\Phi$ is left-invertible (for the state estimation),
it is obvious from $\Phi_I=\pi_I\circ\Phi$ that $\pi_I$ is left-invertible ($\Phi$ is redundant) if and only if $\Phi_I$ is left-invertible.
}
to the assumption made for Corollary~\ref{cor:centralized}.
And, for the linear systems,
i.e., 
when the system \eqref{eq:plant} as well as the function $\Phi$ defined in \eqref{eq:phi} are given as linear,
it becomes the same assumption commonly considered as in \cite{Fawzi14,Chong15ACC,Chanhwa19TAC};
each function
$\Phi_I=\{\Phi_i\}_{i\in I}$ with $| I|=p-2q$
is injective as a linear map.
\end{rem}

Now, let us recall that one of the objectives of this paper is to reduce the computational effort for solving the resilient state estimation.
As the problem is known as combinatorial in nature,
as investigated in
\cite{Pasqualetti13TAC},
the number of cases for choosing and identifying the index set $ I$
is $\binom{p}{q}=\frac{p!}{q!(p-q)!}$,
which rapidly grows
as the number $p$ of measurements increases.
In this regard,
our approach for reducing the complexity is to divide
the set of estimates $\{\hat z_i(t)\}_{i\in[p]}$ into local sets of estimates,
and perform {\it  the attack identification with respect to each local set}.

Let $P\subset[p]$ be an index set for a set of local measurements, and let $\{\hat z_i(t)\}_{i\in P}$ and $\{\Phi_i\}_{i\in P}$ be the corresponding partial estimates and functions, respectively.
For the attack identification only,
requiring the functions $\{\Phi_i\}_{i\in P}$ to satisfy the existing $2q$-redundant observability condition,
i.e.,
requiring that they are left-invertible even when any $2q$ elements of them are removed,
would be restrictive,
since it is hard to expect the system \eqref{eq:plant} observable from a local subset of measurement, let alone the redundancy condition.
In contrast, as seen in Theorem~\ref{thm:identification} and Remark~\ref{rem:red_obs},
the proposed redundancy in Section~\ref{subsec:redundancy}
is less restrictive in the sense that the observability (injectivity) is not required.
Indeed, it can be easily checked from Theorem~\ref{thm:identification} that the attack identification
can be performed for the local estimates $\{\hat z_i(t)\}_{i\in P}$, instead of $\{\hat z_i(t)\}_{i\in [p]}$, as long as the local function $\{\Phi_i\}_{i\in P}$ is $2q$-Lipschitz redundant.
Note that the local function $\Phi_P(x)=\col\{\Phi_i(x)\}_{i\in P}$
is defined to be $k$-(Lipschitz) redundant,
if
the condition of
Definition~\ref{def:redundant} (or \ref{def:Lipschitz_redundant})
holds with respect to the canonical projection
$$\pi_{P, I}: \Phi_P(x)=\col\{\Phi_i(x)\}_{i\in P}\mapsto\Phi_I(x)=\col\{\Phi_i(x)\}_{i\in I}, $$
for each $ I\subset P$ such that $| I| \ge |P|-k$.

Meanwhile, the following example shows that a function that is not redundant
may become redundant,
by taking a projection mapping to keep the redundant part only.

{\it Example:}
Let $x=(r,\theta)\in \XXXX= [1,2]\times [0,\pi/4]$, and let
$$
{\Phi(x) = \begin{bmatrix}
\Phi_1(x)\\\Phi_2(x)\\\Phi_3(x)
\end{bmatrix} = \begin{bmatrix}
\begin{bmatrix}
r\cos \theta\\
r \sin \theta
\end{bmatrix}
\\
\tan\theta\\
\theta
\end{bmatrix}\in\R^{(2,1,1)}.
}$$
The function $\Phi$ is clearly not $2$-redundant,
as the map $\pi_{\{3\}}:\Phi(x)\mapsto \Phi_3(x)=\theta$ is not injective and the value of $r$ cannot be restored from $\theta$.
But, with a projection $T(z)= z/\sqrt{z^\top z}$, $z\in\R^2$,
which removes the information of $r$ from $\Phi_1(x)$, as $T(\Phi_1(x))=[\cos \theta,\sin\theta]^\top$,
the projected function $\Phi'(x)=\col\{T(\Phi_1(x)),\Phi_2(x),\Phi_3(x)\}$ becomes $2$-redundant.

With this observation,
the scheme of local attack identification is presented,
where a set of projections can be used for refining the sets $\{\Phi_i\}$ of functions to be redundant.
Let $P_1, P_2,\dots,P_l\subset [p]$ be local index sets
that cover the set $[p]$, i.e.,
$\cup_{j=1}^{l}P_j=[p]$,
and for each $j=1,\dots,l$
and $i\in P_j$,
let $T_i^j:\R^{n_i}\ra \R^{n_i^j}$, $n_i^j\le n_i$, be a Lipschitz function for the projection
of the estimate $\hat z_i(t)\in\R^{n_i}$,
with respect to $P_j$.

Next, we formulate an equation for the local attack identification;
with
an indexed sequence
$\NN_j:=\{n_i^j\}_{i\in P_j}$,
let
\begin{align}\label{eq:xi}
\begin{split}
&\hat \xi^j(t) := \col \{\hat \xi_i^j(t)\}_{i\in P_j}:= \col\{T_i^j(\hat z_i(t))\}_{i\in P_j}\in\R^{\NN_j},\\
&\Psi^j(x):=\col\{\Psi^j_i(x)\}_{i\in P_j}:=\col\{T_i^j(\Phi_i(x))\}_{i\in P_j}\in\R^{\NN_j},
\end{split}
\end{align}
where we abuse notation and let $\R^{\NN_j}$ denote the space of partitioned vectors whose indices belong to $P_j$.
By \eqref{eq:convergence} and the Lipschitzness of $T_i^j$, note that the local estimate $\hat \xi_i^j(t)$ satisfies
$$\| \hat \xi_i^j(t) - \Psi_i^j(x(t))\| \le L(T_i^j) \delta_i(t).$$
Then,
as the equation
\eqref{eq:static} derived from \eqref{eq:convergence} and \eqref{eq:static_individual},
an equation with respect to the $j$-th estimate $\hat \xi^j(t)$ is established as
\begin{equation}\label{eq:static_local}
\hat \xi^j(t) = \Psi^j(x(t)) + r^j(t)+ e^j(t),
\end{equation}
where $r^j(t)\in\R^{\NN_j}$ and  $e^j(t)\in\R^{\NN_j}$ are the observer error and the effect of attack, respectively,
subject to the constraints
\begin{equation}\label{eq:constraint_local}
\|r^j(t)\| \le \delta^j(t):=\max_{i\in P_j}\{L(T_i^j)\delta_i(t)\},~\,\|e^j(t)\|_0^{\NN_j}\le q.
\end{equation}

Now,
we propose that
the attack identification be performed
for each local estimate
$\hat \xi^j(t)$.
Let the condition
\eqref{eq:inspection} used for the method of Theorem~\ref{thm:identification} be written with respect to \eqref{eq:static_local}, as\footnote{
Since $\hat \xi^j=\col\{\hat \xi_i^j\}_{i\in P_j}$ and $\Psi^j=\col\{\Psi_i^j\}_{i\in P_j}$,
we again abuse notation and let
$\hat \xi_{I_j}^j:=\col\{\hat \xi_i^j\}_{i\in {I_j}}$ and $\Psi_{I_j}^j:=\col\{\Psi_i^j\}_{i\in {I_j}}$, for ${I_j}\subset P_j$.
}
\begin{equation}\label{eq:inspection_local}
d(\hat \xi_{I_j}^j(t),\Psi_{I_j}^j(\XXXX))\le \delta^j(t)
\end{equation}
where the index set ${I_j}\subset P_j$ removes $q$ elements out of $\hat \xi^j(t)=\{\hat \xi_i^j(t)\}_{i\in P_j}$, i.e.,
${I_j}\subset P_j$ is such that $|{I_j}|=|P_j|-q$.

As a result,
the following theorem
presents our approach to the resilient state estimation,
based on the proposed attack identification applied to each local estimate $\hat \xi^j(t)$.
See Fig.~\ref{fig:config} describing the overall configuration of the proposed scheme.

\begin{thm1}\label{thm:local}
Consider the observers  \eqref{eq:observer} under the attack subject to \eqref{eq:unattacked}, and the local estimates
defined as \eqref{eq:xi}.
Assume that the function $\Psi^j$ is $2q$-Lipschitz redundant, $\forall j=1,\dots,l$.
\begin{enumerate}
\item\label{item:thm2.1} Let ${I_j}\subset P_j$ such that $|{I_j}| = |P_j| - q$ be an index set. If \eqref{eq:inspection_local} is violated, then ${I_j}\!\setminus I_0 \!\neq \emptyset$, i.e., there is an corrupted estimate $\hat \xi_i^j(t)$ such that $i\in[p]\setminus I_0$.
If \eqref{eq:inspection_local} holds, then
\begin{equation}\label{eq:attack_free_local}
\|\hat \xi_{I_j}^j(t) -  \Psi_{I_j}^j(x(t)) \|\le (2 M_{\Psi^j} + 1)\delta^j(t)
\end{equation}
where $M_{\Psi^j}$ is a constant given from Definition~\ref{def:Lipschitz_redundant},
from the $2q$-Lipschitz redundancy of the function $\Psi^j$.
\item\label{item:thm2.2}
In addition,
assuming that
$\col\{\Psi^j(\cdot)\}_{j=1}^{l}$ is an injective immersion on $\XXXX$,
there exist
index sets
$\{I_j\}_{j=1}^{l}$
with $|I_j|\!=\!|P_j|-q$, $\forall j$,
and
a Lipschitz function $\psi$, such that 
\begin{equation}\label{eq:recover_state_local}
\left\|\psi\left(\col\{\hat \xi_{I_j}^j(t)\}_{j=1}^{l}\right) -  x(t) \right\|\le L(\psi)\delta'(t),~~\forall t\ge 0,
\end{equation}
holds,
where
$\delta'(t):= \max_{j}\{(2 M_{\Psi^j} + 1)\delta^j(t)\}$.
\hfill$\square$
\end{enumerate}
\end{thm1}

Theorem~\ref{thm:local}.\ref{item:thm2.1} means that,
for each set
$\{\hat \xi_i^j(t)\}_{i\in P_j}$ of local estimates,
the un-corrupted elements can be identified,
when the function $\Psi^j$ is redundant. Theorem~\ref{thm:local}.\ref{item:thm2.2} means that
the information of the state $x(t)$ as well as that of $\{\Psi^j(x(t))\}_{j=1}^l$ can be restored from the locally identified estimates, when the collection  $\col\{\Psi^j(x)\}_{j=1}^{l}$ of functions is left-invertible.

{\it Proof of Theorem~\ref{thm:local}:}
1)
Let $e^j(t) = \col\{e_i^j(t)\}_{i\in P_j}\in\R^{\NN_j}$.
By applying Theorem~\ref{thm:identification}.\ref{item:thm1.1} with respect to the equation \eqref{eq:static_local} with \eqref{eq:constraint_local},
it follows that \eqref{eq:inspection_local} implies \eqref{eq:attack_free_local}, as the same as \eqref{eq:inspection} implies \eqref{eq:attack_free}.
And, conversely, it also follows that the negation of \eqref{eq:inspection_local} implies that $e_{I_j}^j(t)=\col\{e_i^j(t)\}_{i\in I_j}\neq 0$,
which implies $e_i^j(t)\neq 0$ for some $i\in I_j$, i.e., $i\not\in I_0$.
2)
From Theorem~\ref{thm:identification}.\ref{item:thm1.2}, it can be checked that
for $j=1,\dots,l$,
there exist $I_j \subset P_j$ with $|I_j|= |P_j|-q$ satisfying \eqref{eq:inspection_local} and \eqref{eq:attack_free_local},
and a Lipschitz function $\psi_{I_j}$
satisfying $\psi_{I_j}(\Psi_{I_j}^j(x)) = \Psi^j(x)$, $\forall x\in\XXXX$.
Since the map $\col\{\Psi^j(\cdot)\}_{j=1}^{l}$ is an injective immersion on $\XXXX$,
by \cite[Proposition~1]{Kim19TAC} (see Remark~\ref{rem:immersion}),
there exists a Lipschitz map $\Psi^{-1}$ such that $\Psi^{-1}(\col\{\Psi^j(x)\}_{j=1}^{l})=x$, $\forall x\in\XXXX$.
Define
$\psi(\cdot):= \Psi^{-1}(\col\{\psi_{I_j}(\cdot)\}_{j=1}^{l})$.
Then,
since
$\|\col\{\hat \xi_{I_j}^{j}(x)\}_{j=1}^{l} -  \col\{\Psi_{I_j}^{j}(x(t))\}_{j=1}^{l}\|\le \delta'(t)$ holds by \eqref{eq:attack_free_local},
it is clear that
\eqref{eq:recover_state_local} holds.
The proof is completed.
\hfill$\blacksquare$

\begin{rem}\label{rem:complexity}
While the number of cases that the conventional attack identification considered in Corollary~\ref{cor:centralized} is $\binom{p}{q}$,
the proposed scheme by local identification requires that $\sum_{j=1}^{l}\binom{|P_j|}{q}$ cases are considered, since the number of the cases for the $j$-th local set $\{\hat \xi_i^j(t)\}_{i\in P_j}$ is $\binom{|P_j|}{q}$.
Thus, the larger number $p$ of the sensors and the smaller number of the elements for each group $P_j$ are considered,
the more benefit the proposed scheme takes, in terms of computational efforts and consumed resources for running the algorithm.
\end{rem}

With the introduced redundancy notion,
our scheme of local attack identification has been proposed,
for the purpose of reducing the complexity of resilient state estimation.
Then,
a follow-up question would be:
How can \eqref{eq:static} be divided into the local equations \eqref{eq:static_local} in practice,
i.e.,
how can the local groups of partial estimates, which should keep the $2q$-redundancy, be found?
To answer this question,
in the next subsection,
we introduce a class of systems
for which the local groups $\{P_j\}_{j=1}^l$ and the projections $\{T_i^j\}_{i\in P_j}$ can be found, in a particularly efficient and constructive way.
It will be also be seen that the introduced class includes all linear systems.

\subsection{Constructive Design for a Class of Systems}\label{subsec:cases}

Let us get back to \eqref{eq:static} formulated from the design of observers \eqref{eq:observer},
in which the function $\{\Phi_i\}_{i\in [p]}$ has been defined in \eqref{eq:phi}, from the vector field $f$ and the individual output functions $\{h_i\}_{i\in [p]}$ of the system \eqref{eq:plant}.

The assumption made in this subsection for
more constructive design is that,
for each $i\in[p]$, the function
$\Phi_i$
can be decomposed with respect to a unified coordinate function of the domain $\XXXX$;
there exist diffeomorphisms $T_x$ on $\XXXX$, and $T_i$
on $\Phi_i(\XXXX)\subset\R^{n_i}$,
so that with respect to the sub-states
\begin{equation}\label{eq:xxx}
\xxx= \col\{\xxx^j\}_{j=1}^{l}:=T_x(x)\in\R^n,
\end{equation}
the map
$\phi_i(\xxx):=
T_i\circ\Phi_i\circ T_x^{-1}(\xxx)\in\R^{n_i}
$
is decomposed as
\begin{equation}\label{eq:decomposition_phi}
\phi_i(\xxx)
= \col\{\phi_i^j(\xxx^j)\}_{j\in N_i}
\end{equation}
with some functions
$\{\phi_i^j\}_{j\in N_i}$,
where we define
\begin{equation}\label{eq:N_i}
N_i := \left\{j=1,\dots,l: \frac{\partial \phi_i}{\partial \xxx^j}\not\equiv 0\right\}, 
\end{equation}
i.e.,
the sub-states
$\{\xxx^j\}_{j\in N_i}$
determine the value of $\phi_i(\xxx)$, and $\phi_i(\xxx)$ is constant with respect to the rest $\{\xxx^j\}_{j\not\in N_i}$.
Regarding the dimensions,
for each $j=1,\dots, l$,
let $\xxx^j\in\R^{n^j}$ 
and
$\phi_i^j(\xxx^j)\in\R^{n_i^j}$,
with some
$n^j\in\N$ and
$n_i^j\in\N$,
respectively,
so that $\sum_{j=1}^{l}n^j=n$
and $\sum_{j\in N_i}n_i^j = n_i$.

Now,
the sets $\{P_j\}_{j=1}^{l}$ and the projections $\{T_i^j\}_{i\in P_j}$,
which yields the local equations \eqref{eq:static_local} by \eqref{eq:xi},
are simply found as
\begin{align}\label{eq:P_j}
\begin{split}
P_j&:= \{i\in [p]:j\in N_i\},\quad j=1,\dots, l,\\
T_i^j&:= \pi_{N_i, j}\circ T_i,~ \text{where}~ \pi_{N_i, j}: \col\{\phi_i^j(\xxx^j)\}_{j\in N_i}\mapsto \phi_i^j(\xxx^j).
\end{split}
\end{align}

\begin{figure}[!h]
	\centering
		\includegraphics[width=0.4\textwidth]{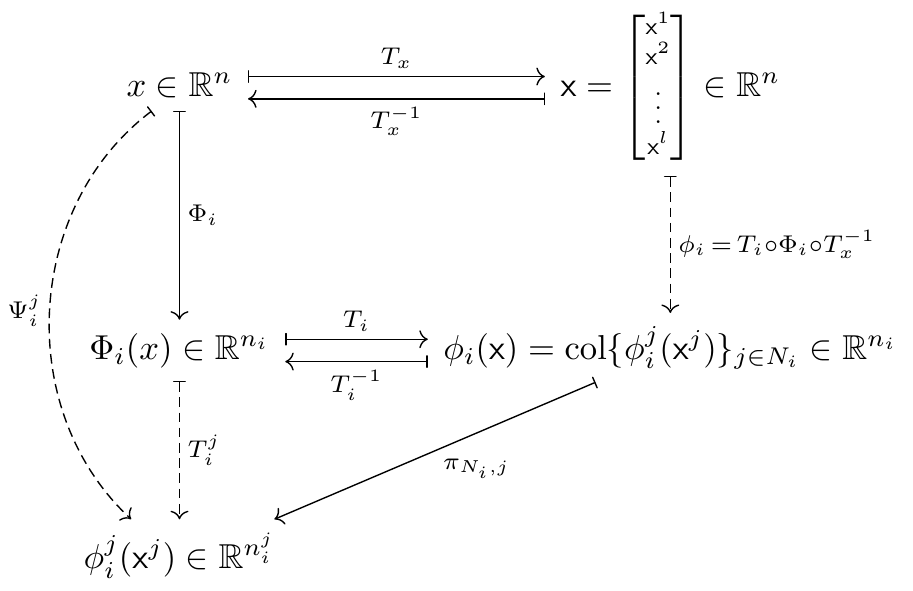}
	\caption{
	Commutative diagram for the functions $T_i^j$ and $\Psi_i^j$.	}
	\label{fig:commutative3}
\end{figure}

In summary,
the local equations \eqref{eq:static_local} can be established from \eqref{eq:static} in a constructive way,
from the knowledge of the coordinate changes $T_x$ and $\{T_i\}_{i\in[p]}$.
See Fig.~\ref{fig:commutative3} showing a commutative diagram
for the projection maps $\{T_i^j\}$ and the local function $\Psi^j=\col\{\Psi_i^j\}_{i\in P_j}$ defined from the condition \eqref{eq:decomposition_phi}.

Technically, we first define the function $T_i^j$ on the domain $\Phi_i(\XXXX)$ on which the map $T_i$ is defined, so that $T_i^j(\Phi_i(x)):= \phi_i^j(\xxx^j)$ for $x\in\XXXX$, and then consider Lipschitz extension so that the map $T_i^j$ becomes a Lipschitz function defined on $\R^{n_i}$.

Then,
we have the following theorem;
if the function $\Phi$ of \eqref{eq:static} can be decomposed as \eqref{eq:decomposition_phi},
then the conditions of Theorem~\ref{thm:local} are satisfied, i.e.,
the proposed approach of local identification is applicable,
as long as the given resilient state estimation problem \eqref{eq:static} is solvable.

\begin{thm1}\label{thm:decomposed}
Assume that
\eqref{eq:decomposition_phi} holds,
and
consider the local equations \eqref{eq:static_local}
derived from
\eqref{eq:static}, \eqref{eq:P_j}, and \eqref{eq:xi}.
Then,
\begin{enumerate}
\item the function $\Psi^j$ is $2q$-Lipschitz redundant for all
$j$,
if and only if the map $\Phi$ is $2q$-Lipschitz redundant.
\item the function $\col\{\Psi^j(\cdot)\}_{j=1}^{l}$ is an injective immersion on $\XXXX$, if and only if
$\Phi$ is an injective immersion on $\XXXX$.\hfill$\square$
\end{enumerate}
\end{thm1}

We recall that
the conditions for $\Phi$ in Theorem~\ref{thm:decomposed} have been considered as necessary condition for the resilient state estimation, as discussed in Remarks~\ref{rem:immersion} and \ref{rem:red_obs}.

{\it Proof:}
1)
Define
$T_I(\col\{z_i\}_{i\in I}):= \col\{T_i(z_i)\}_{i\in I}$
for $I\subset [p]$,
where
$z_i\in \Phi_i(\XXXX)$,
which is clearly a diffeomorphism on the image of $\col\{\Phi_i(\cdot)\}_{i\in I}$.
Suppose that $\Phi$ is $2q$-Lipschitz redundant,
choose $\jmath\in\{1,\dots,l\}$,
and let a function $\Psi^\jmath_{I_\jmath}(x)=\col\{\Psi_i^\jmath(x)\}_{i\in I_\jmath}$ be given, with $I_\jmath \subset P_\jmath$ such that
$|I_\jmath|\ge |P_\jmath|-2q$.
Since $I' := I_\jmath\cup ([p]\setminus P_\jmath)$ satisfies $|I'|\ge p-2q$,
there exists a Lipschitz map $\pi_{I'}^{-1}$ such that $\pi_{I'}^{-1}(\Phi_{I'}(x))=\Phi(x)$, $\forall x\in\XXXX$.
Then,
with
$\Gamma:=T_{[p]}\circ\pi_{I'}^{-1}\circ T_{I'}^{-1}$, it follows that
\begin{align*}
\Gamma(\col\{\col\{\phi_i^j(\xxx^j)\}_{j\in N_i}\}_{i\in I'})
&= \Gamma(\col \{T_i(\Phi_i(x))\}_{i\in I'})\\
&=T_{[p]}(\Phi(x))\\
&=\col\{\col\{\phi_i^j(\xxx^j)\}_{j\in N_i}\}_{i\in [p]}
\end{align*}
holds by \eqref{eq:xxx} and
\eqref{eq:decomposition_phi},
for all $\col\{\xxx^j\}_{j=1}^{l}\in T_x(\XXXX)$.
Thus, by taking the components that correspond to the index $j=\jmath$ only,
by $j\in N_i\Leftrightarrow i\in P_j$,
we have a Lipschitz function
$\Gamma^\jmath$ that satisfies
$
\Gamma^\jmath(\col\{\phi_i^\jmath(\xxx^\jmath)\}_{i\in I'\cap P_\jmath})
=\col\{\phi_i^\jmath(\xxx^\jmath)\}_{i\in P_j}
$, $\forall \xxx^\jmath$.
Then,
since $I'\cap P_\jmath=I_\jmath$ and $\Psi_i^\jmath(x)=T_i^\jmath(\Phi_i(x))=\phi_i^\jmath(\xxx^\jmath)$,
we have $\Gamma^\jmath(\Psi_{I_\jmath}^\jmath(x)) = \Psi^\jmath(x)$, $\forall x\in\XXXX$. Hence, $\Psi^j$ is $2q$-Lipschitz redundant, $\forall j$.
Conversely, suppose that the maps $\{\Psi^j\}_{j=1}^{l}$ are $2q$-Lipschitz redundant,
and let $I\subset[p]$ be given, with $|I|\ge p-2q$.
Let $I_j := I\cap P_j$ for each $j$, so that $|I_j|\ge |P_j|-2q$. Then, there exist Lipschitz functions $\{\Gamma^j\}_{j=1}^{l}$ such that $\Gamma^j(\Psi_{I_j}^j(x))=\Psi^j(x)$, $\forall x\in\XXXX$, $\forall j$,
so that we can define
a map $\Gamma'$ such that
$\Gamma'(\col\{\Psi_{I_j}^j(x)\}_{j=1}^l)= \col\{\Psi^j(x)\}_{j=1}^{l}$, $\forall x\in\XXXX$.
Since $\{\Psi_{I_j}^j(x)\}_{j=1}^{l}=
\{\{\Psi_i^j(x)\}_{i\in I_j}\}_{j=1}^l
$ and since $i\in I_j\Leftrightarrow (j\in N_i)\wedge(i\in I) $, we can also define $\Gamma''$ such that
\begin{equation*}
\Gamma'' (\col\{\col\{\Psi_i^j(x)\}_{j\in N_i}\}_{i\in I})
=  \col\{\col\{\Psi_i^j(x)\}_{j\in N_i}\}_{i\in [p]},
\end{equation*}
$\forall x\in\XXXX$.
By
$\col\{\Psi_i^j(x)\}_{j\in N_i}=T_i(\Phi_i(x))$
and
the definitions of $T_I$ and $T_{[p]}$,
we have
$\Gamma''(T_I(\Phi_I(x)))=T_{[p]}(\Phi(x))$, $\forall x\in\XXXX$.
Then, the map $\pi_I^{-1}:= T_{[p]}^{-1}\circ\Gamma'' \circ T_I$ for each $I$ proves that
$\Phi$ is $2q$-Lipschitz redundant.
2) Since
$T_{[p]}$ is a diffeomorphism,
$\Phi(x)$ is an injective immersion if and only if
the map
$T_{[p]}\circ \Phi(x)=\col\{\col\{\Psi_i^j(x)\}_{j\in N_i}\}_{i\in [p]}$ is an injective immersion, which is equivalent to the condition that $\col\{\Psi^j(x)\}_{j=1}^{l}=\col\{\col\{\Psi_i^j(x)\}_{i\in P_j}\}_{j=1}^{l}$ is an injective immersion.
\hfill $\blacksquare$

\begin{rem}
The $j$-th local equation \eqref{eq:static_local} with the function $\Psi^j(x)=\{\phi_i^j(\xxx^j)\}_{i\in P_j}$ is formulated with respect to the $j$-th sub-state $\xxx^j$ only.
Then,
continuing the discussion of Remark~\ref{rem:complexity},
we note from \eqref{eq:N_i} that
the cardinality of $P_j$,
which determines the complexity $\binom{|P_j|}{q}$ of the $j$-th local problem, is given by the number of the functions $\phi_i$ (the number of the functions $\Phi_i$, equivalently) of which $\xxx^l$ is an independent variable.
\end{rem}

Finally, the following proposition states that
the condition \eqref{eq:decomposition_phi} can be satisfied whenever the system \eqref{eq:plant} is linear, so that
the proposed method with the construction \eqref{eq:P_j} is applicable.

\begin{prop}\label{prop:linear}
If
the vector field $f$ and the function $h$ of the system \eqref{eq:plant} are linear,
then
there exist invertible linear maps $T_x$ and $\{T_i\}_{i\in[p]}$ such that the condition \eqref{eq:decomposition_phi} is satisfied.
\end{prop}

{\it Proof:}
Let $A\in\R^{n\times n}$ and $C_i\in\R^{1\times n}$ be the matrix representations of the linear maps $f$ and $h_i$, $i\in[p]$, respectively.
Let the characteristic polynomial $\gamma_A(s)=\det(sI_n-A)$ of $A$ be factorized as $\gamma_A(s)=\Pi_{j=1}^{l} \gamma^j(s)$,
where for each $j_1\neq j_2$,
$\gamma^{j_1}(s)$ and $\gamma^{j_2}(s)$ are relatively prime.
We transform $A$  into a block diagonal matrix with an invertible linear map $T_x$,
as $\mathrm{diag}\{\AAA^j\}_{j=1}^{l}:= T_x A T_x^{-1}$,
where
$\AAA^j\in\R^{n^j\times n^j}$, $\sum_{j=1}^{l}n^j=n$, is such that
$\det(sI_{n^j}-\AAA^j)=\gamma^j(s)$.
Next, let $
\col\{\xxx^j\}_{j=1}^{l}:=T_x \cdot x$ with $\xxx^j\in\R^{n^j}$, and for $i\in[p]$, let $[\CCC_i^1,\CCC_i^2,\dots,\CCC_i^l]:=C_iT_x^{-1}$ with $\CCC_i^j\in \R^{1\times n^j}$.
Here, note that
\begin{align}\label{eq:CA}
\begin{split}
&\Phi_i(x) = 
\col\{C_i A^{k-1}x\}_{k=1}^{n}
\\ &={\begin{bmatrix}
\CCC_i^1 & \CCC_i^2& \cdots& \CCC_i^l\\
\CCC_i^1\AAA^1 & \CCC_i^2\AAA^2& \cdots& \CCC_i^l\AAA^l\\
\vdots & \vdots& \ddots& \vdots\\
\CCC_i^1(\AAA^1)^{n-1} & \CCC_i^2(\AAA^2)^{n-1}& \cdots& \CCC_i^l(\AAA^l)^{n-1}\\
\end{bmatrix}\cdot\begin{bmatrix}
\xxx^1\\\xxx^2\\\vdots\\\xxx^l
\end{bmatrix}},
\end{split}
\end{align}
from \eqref{eq:phi}. Now, we claim that
if $\Phi_i(x)=0$, then $\CCC_i^j(\AAA^j)^{k-1} \xxx^j=0$, $\forall j=1,...,l$ and $k=1,..., n$.
Suppose that $\Phi_i( x)=0$, and choose $\jmath\in\{1,...,l\}$ and $\kappa\in\{1,...,\kappa\}$.
Then it follows that $C_i\gamma(A)x=0$ for any polynomial $\gamma$,
where we
define $\gamma(A):= \sum_{k=1}^{n}c_k A^k$ for $\gamma(s)=\sum_{k=1}^{n}c_k s^k$.
It is followed by
$\sum_{j=1}^{l}\CCC_i^j \gamma(\AAA^j)\xxx^j=0$, by the diagonal structure of $\mathrm{diag}\{\AAA^j\}$.
Since two polynomials $\gamma^\jmath(s)$ and $\gamma'(s):=\Pi_{j\in \{1,...,l\}\setminus\{\jmath\}}\gamma^j(s)$ are relatively prime, there exist two polynomials $\zeta(s)$ and $\zeta'(s)$ such that 
$\zeta(s)\gamma^\jmath(s) + \zeta'(s)\gamma'(s)= s^{\kappa-1}$,
so we choose $\gamma$ as $\gamma(s):= s^{\kappa-1}-\zeta(s)\gamma^\jmath(s)
= \zeta'(s)\gamma'(s)
$.
Since $\gamma^j(\AAA^j)=0$, $\forall j=1,...,l$,
it follows that
$\gamma(\AAA^\jmath)= (\AAA^\jmath)^{\kappa-1}$,
and
$\gamma(\AAA^j)=0$ if $j\neq \jmath$.
Thus, from
$\sum_{j=1}^{l}\CCC_i^j \gamma(\AAA^j)\xxx^j=0$,
we have $\CCC_i^\jmath(\AAA^\jmath)^{\kappa-1}\xxx^{\jmath}=0$, so the claim is proved.
From \eqref{eq:CA}, define
$\Phi_i(x)=:\sum_{j=1}^{l}\OO_i^j \xxx^j$,
and $N_i:=\{j\in\{1,...,l\}:\OO_i^j\neq 0\}$, which is equivalent to \eqref{eq:N_i}.
And, for each $j\in N_i$,
let $\{\tau_1^j,...,\tau_{n_i^j}^j\}$, $n_i^j\in\N$, be column vectors that consist of a basis of the image space of $\OO_i^j$,
with which we define
$\OO_i^j\xxx^j=:[\tau_1^j,...,\tau_{n_i^j}^{j}]\cdot\phi_i^j(\xxx^j)$.
Then, the claim implies that the intersection of the image spaces of $\OO_i^j$ is equal to zero,
so that with $\sum_{j\in N_i} {n_i^j}=\mathrm{rank} (\Phi_i) = n_i$,
the collection $\{\tau_k^j:k=1,...,n_i^j,\,j\in N_i\}$ is a basis of $\R^{n_i}$.
Hence,
with
the inverse matrix
$T_i:= [\tau_1^1,...,\tau_{n_i^1}^1,\tau_1^2,...,\tau_{n_{i}^l}^{l}]^{-1}$, we have
\begin{align*}
T_i\cdot \Phi_i(x) &=  T_i\cdot \begin{bmatrix}
\OO_i^1,\dots,\OO_i^l
\end{bmatrix}
\cdot T_x^{-1} \cdot \col\{\xxx^j\}_{j=1}^{l}\\&= \col\{\phi_i^j(\xxx^j)\}_{j\in N_i},
\end{align*}
i.e., we have \eqref{eq:decomposition_phi}.
The proof is completed.\hfill$\blacksquare$

When the system \eqref{eq:plant} is linear,
the problem can be divided as many times as the number of distinct poles of \eqref{eq:plant},
because,
the number $l$ has been found as
the number of relatively prime factors of the characteristic polynomial of
the map
$f$.
The number of cases $\binom{|P_j|}{q}$ of the $j$-th local problem is determined by 
$P_j = \{i: \CCC_i^j\neq 0 \}$,
i.e., the number of individual measurements whose output matrix $\CCC_i^j$ with respect to the sub-state $\xxx^j$ is non-zero.
This is because $\Phi_i(x) = \sum_{j=1}^{l}\OO_i^j\xxx^j$ in \eqref{eq:CA}, and $\OO_i^j =0$ if and only if $\CCC_i^j=0$.

\section{Numerical Example}\label{sec:simulation}

Consider a numerical example of the system \eqref{eq:plant}, given as
\begin{equation*}
{\begin{bmatrix}
\dot x_1\\
\dot x_2\\
\dot x_3
\end{bmatrix}
=
\begin{bmatrix}
-x_1+ \frac{1}{2}x_3^2-x_2 x_3 \cos x_2 \\
-x_2\\
-x_2 \cos x_2
\end{bmatrix}
+ \begin{bmatrix}
x_3 + x_3\cos x_2\\
1\\
1+\cos x_2
\end{bmatrix}u
}
\end{equation*}
with $u(t)= 0.25\sin(0.2\pi t)$, where
the state $x(t)$ remains in the set $\XXXX=\{x\in\R^3: \|x \|\le 0.5\}$ when its initial value is given sufficiently small.
For the outputs \eqref{eq:plant_output}, let there be $20$ measurements
with the functions $h=\{h_i\}_{i=1}^{20}$ given as
\begin{equation*}
h_i(x) = \begin{cases}
x_1 - \frac{1}{2} x_3^2 + \frac{i}{10} x_2, ~~\quad i=1,\dots, 10,\\
\frac{1}{2}x_3 - \frac{1}{2}\sin x_2,~~\qquad i=11,\dots, 20,
\end{cases}
\end{equation*}
and the noise $v(t)\in\R^{20}$ be given such that $\| v(t)\| \le 0.01$.

In this example,
the partial information $\Phi_i(x)$ of the state $x$
obtained from each individual output $y_i$ is no more than $h_i(x)$, i.e., $\Phi_i=h_i$, $\forall i$,
as it can be computed from \eqref{eq:phi} that $L_f h_i= -h_i$ for $i\le 10$, and
$L_f h_i = 0$ for $i\ge 11$.

Then,
it can be verified that
the state $x(t)$ can be restored from any $12$ elements of the measurements $\{y_i(t)\}_{i=1}^{20}$,
so that the attack identification and
the resilient state estimation is possible when up to $4$ measurements are compromised.
If the attack identification is performed with all the measurements, then the number of cases is equal to $\binom{20}{4}=4845$.

But,
the attack identification can also be performed with the local measurements $\{y_i(t)\}_{i=1}^{10}$ and $\{y_i(t)\}_{i=11}^{20}$, separately, although the state $x(t)$ is not observable from either of them.
This is because
both
the collections $\{\Phi_i\}_{i=1}^{10}$ and $\{\Phi_i\}_{i=11}^{20}$ of functions are $8$-redundant, and $8$-Lipschitz redundant on any compact set in $\R^3$, as well;
for example,
removing any $8$ elements from $\{\Phi_i\}_{i=1}^{10}$ and
given any $\Phi_{i_1}(x)$ and $\Phi_{i_2}(x)$ with $1\le i_1 < i_2 \le 10$, both the information of $x_1-\frac{1}{2}x_3^2$ and $x_2$ can be obtained from the combination so that the information of $\{\Phi_i(x)\}_{i=1}^{10}$ can be restored.
As a result, the number of cases for the attack identification is reduced to $2\times \binom{10}{4}=420$.

Regarding the criterion \eqref{eq:inspection_local} for the local inspection for this simple example,
we have $\delta^j(t)\equiv 0.01$,
$\hat \xi_{I_j}^j(t) = \{y_i(t)\}_{i\in I_j}$,
and $\Psi_{I_j}^j(\XXXX)=\Phi_{I_j}(\XXXX)$,
where $I_j \subset \{1,\dots,10\}$ for $j=1$ and $I_j\subset\{11,\dots,20\}$ for $j=2$,
with $|I_j|=6$.
For each case of $|I_j|=6$ with $j=1$,
it satisfies
$\Phi(x) = \OO_{I_j}\cdot[x_1-\frac{1}{2}x_3^2, x_2]^\top$ with some matrix $\OO_{I_j}\in\R^{6\times 2}$,
and it means the set $\Phi_{I_j}(\XXXX)$ is included in the image space of $\OO_{I_j}$.
So we simply check if
\begin{equation}
\left\| \col\{y_i(t)\}_{i\in I_j} - \OO_{I_j}\OO_{I_j}^\dagger\cdot  \col\{y_i(t)\}_{i\in I_j} \right\|_2 \le 0.01\times\sqrt{6},
\label{eq:pseudo}
\end{equation}
where $\|\cdot\|_2$ is the Euclidean norm, and $\OO_{I_j}^\dagger$ is the Moore-Penrose pseudo inverse of $\OO_{I_j}$,
so that the violation of \eqref{eq:pseudo} implies that \eqref{eq:inspection_local} is violated.
For the cases of $j=2$,
we use that
$\Phi_{I_j}(\XXXX)$ is included in the image of $[1,1,\dots,1]^\top\in\R^{6\times 1}$.

\begin{figure}[!t]
	\centering
	\subfigure[]{
		\includegraphics[width=0.35\textwidth]{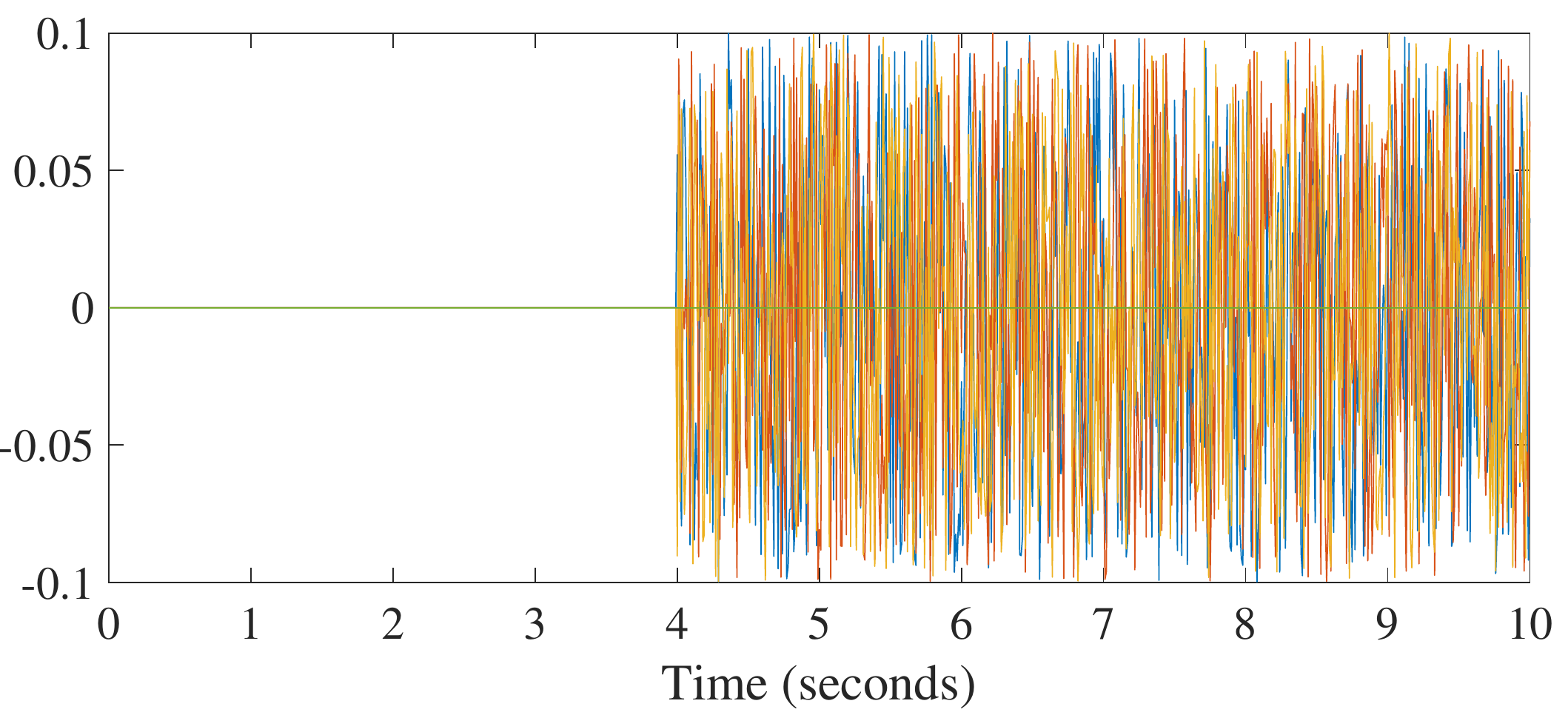}\label{fig:attack}}
	\subfigure[]{
		\includegraphics[width=.35\textwidth]{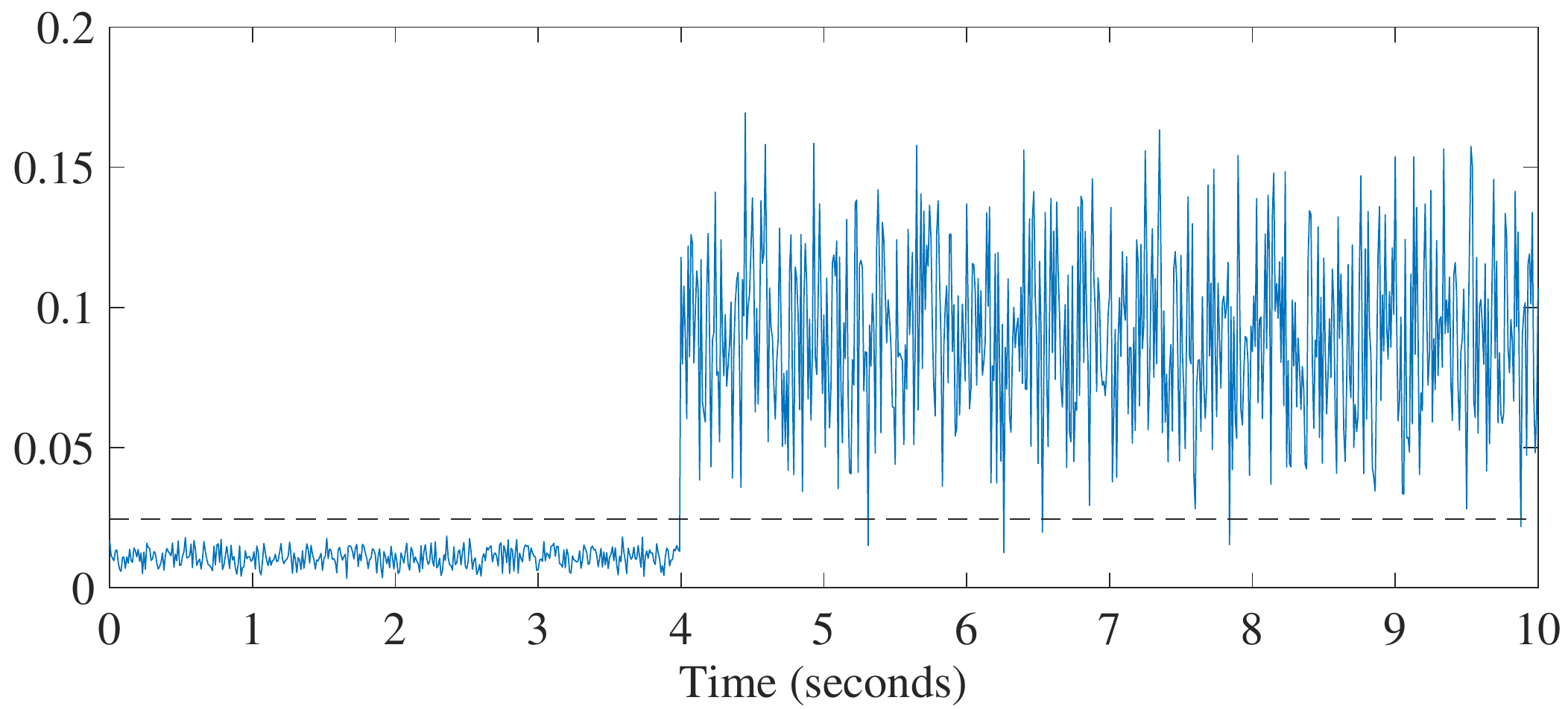}\label{fig:residual1}}
		\subfigure[]{
				\includegraphics[width=.35\textwidth]{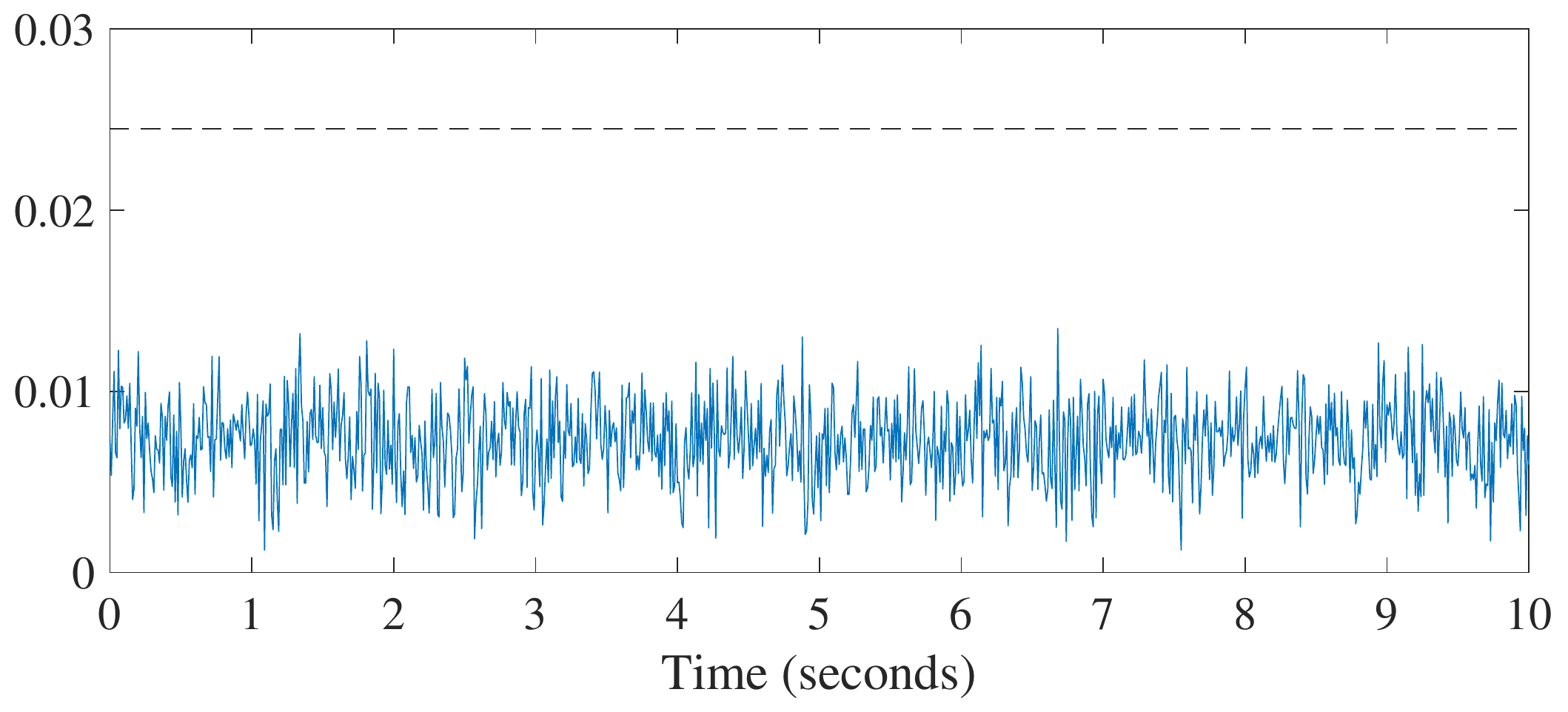}\label{fig:residual2}}
				\subfigure[]{
						\includegraphics[width=.35\textwidth]{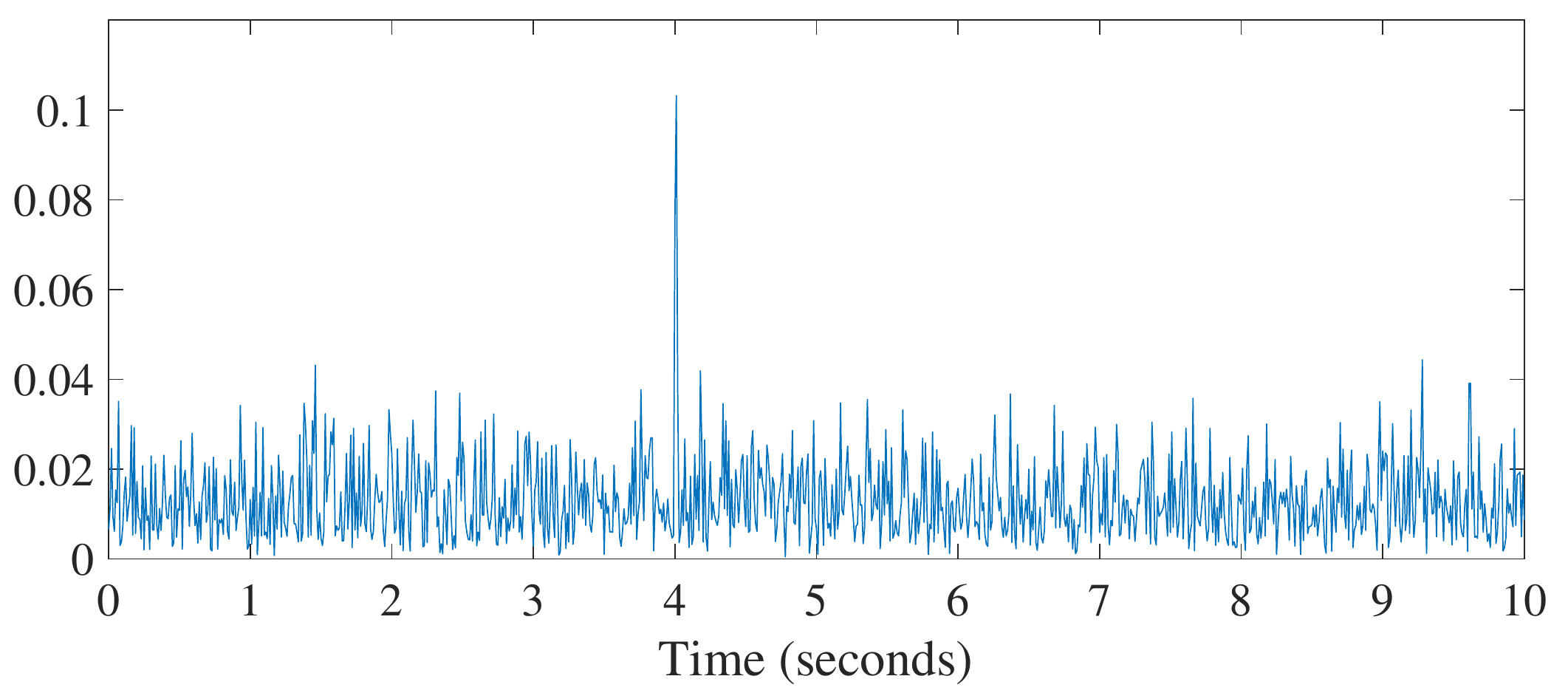}\label{fig:estimate}}
	\caption{
	\small\sf
	Simulation results.
	(a) Injected attack signal $a_i(t)$, $i=1,2,3,4$.
	(b) Distance function $\|\{y_i(t)\}_{i\in I_1} - \OO_{I_1}\OO_{I_1}^\dagger \{y_i(t)\}_{i\in I_1}\|_2$ for monitoring the set $\{ y_i(t)\}_{i\in I_1}$, $I_1=\{1,2,\dots,6\}$, and its threshold
	(black dash-dot).
	(c) Distance function $\|\{y_i(t)\}_{i\in I_1} - \OO_{I_1}\OO_{I_1}^\dagger \{y_i(t)\}_{i\in I_1}\|_2$ for the measurements $\{ y_i(t)\}_{i\in I_1}$, $I_1=\{5,6,\dots,10\}$, and its threshold.
	(d) Estimation error from the state $x(t)$.
	}
	\label{fig:sim}
\end{figure}

Fig.~\ref{fig:sim} shows the simulation results.
Fig.~\ref{fig:attack} depicts the attack scenario; a square wave is injected in the first to the fourth sensors, from the time $t\ge 4$.
Next,
Fig.~\ref{fig:residual1} shows the distance function $\|\{y_i(t)\}_{i\in I_1} - \OO_{I_1}\OO_{I_1}^\dagger \{y_i(t)\}_{i\in I_1}\|_2$ for monitoring the collection $\{ y_i(t)\}_{i\in I_1}$ with its threshold,
where $I_1=\{1,2,\dots,6\}$.
It can be seen that the presence of attack is detected right after it is injected.
And, Fig.~\ref{fig:residual2} shows the same for the index set $I_1 = \{5,6,\dots,10\}$. As it does not contain a corrupted measurement, the signal does not exceed the threshold.
As soon as the attack is detected from a measurement collection,
the monitor switches the index set and
searches out a set of un-corrupted set of measurements,
by inspecting the other combinations and comparing the corresponding distance and the threshold.
For the switching mechanism, we follow the method of \cite{Kim19TAC}.
With the set of identified un-corrupted measurements, the correct estimate of the state $x(t)$ is computed, using left-inverses of the functions. 
Finally, Fig.~\ref{fig:estimate} shows that an accurate estimation for the state $x(t)$ is obtained, except at the moment $t=4$ when the attack is detected and the attack identification is performed.

\section{Conclusion}\label{sec:conclu}
We have proposed an approach for reducing the complexity of resilient state estimation problem for uniformly observable systems,
by performing the attack identification with respect to local groups of projected partial state estimates.
We have introduced a notion of redundancy for a collection of nonlinear functions, and suggested that the attack identification is possible even when the state cannot be reconstructed and is not observable from a set of local measurements, as long as they are redundant.
The sensor groups and the projections have been be found in a constructive manner,
especially
when the partial state information from each individual measurement can be decomposed with respect to a unified coordinate function of the state.
This decomposition is possible for all linear systems.
Future work may consider necessary conditions or sufficient conditions under which the proposed nonlinear decomposition is allowed.

\begin{IEEEbiography}[{\includegraphics[width=1in,height=1.25in,clip]{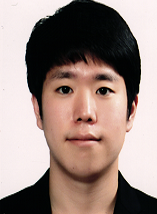}}]{Junsoo Kim}
	received the B.S. degrees in electrical engineering and mathematical sciences in 2014, and the M.S. and Ph.D. degrees in electrical engineering in 2020, from Seoul National University, South Korea, respectively.
	He held the Postdoc position
	at KTH Royal Institute of Technology, Sweden till 2022.
	He is currently an Assistant Professor at the Department of Electrical and Information Engineering, Seoul National University of Science and Technology, South Korea.	
	His research interests include security problems in networked control systems and encrypted control systems.
\end{IEEEbiography}

\begin{IEEEbiography}[{\includegraphics[width=1in,height=1.25in,clip]{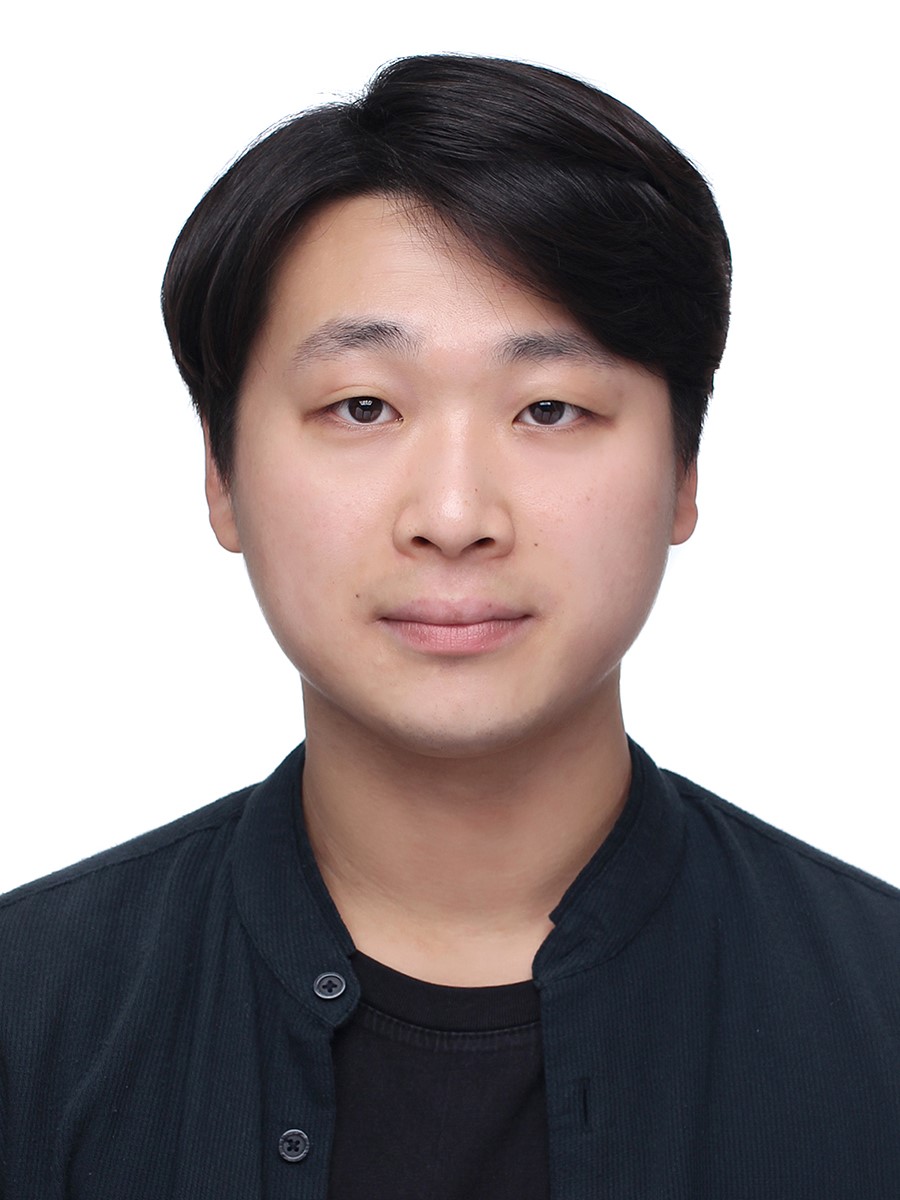}}]{Jin Gyu Lee}
	received his B.S. and Ph.D. degrees from the Department of Electrical Engineering and Computer Science, Seoul National University, Korea, in 2013 and 2019 respectively. He held the post-doc position at Control Group, Department of Engineering, University of Cambridge, United Kingdom till 2021. Then, he held another post-doc position at Control and Power Group, Department of Electrical and Electronic Engineering, Imperial College London, United Kingdom at 2022. He is currently an ISFP (Inria Starting Faculty Position) researcher in Valse team, Inria, France. His research interests include multi-agent systems, observer design, security of cyber-physical systems, nonlinear systems, adaptive control, and neuronal networks.
\end{IEEEbiography}

\begin{IEEEbiography}[{\includegraphics[width=1in,height=1.25in,clip]{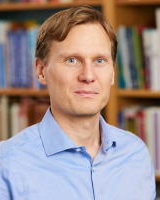}}]{Henrik Sandberg}
is Professor at the Division of Decision and Control Systems, KTH Royal Institute of Technology, Stockholm, Sweden. He received the M.Sc. degree in engineering physics and the Ph.D. degree in automatic control from Lund University, Lund, Sweden, in 1999 and 2004, respectively. From 2005 to 2007, he was a Post-Doctoral Scholar at the California Institute of Technology, Pasadena, USA. In 2013, he was a visiting scholar at the Laboratory for Information and Decision Systems (LIDS) at MIT, Cambridge, USA. He has also held visiting appointments at the Australian National University and the University of Melbourne, Australia. His current research interests include security of cyber-physical systems, power systems, model reduction, and fundamental limitations in control. Dr. Sandberg was a recipient of the Best Student Paper Award from the IEEE Conference on Decision and Control in 2004, an Ingvar Carlsson Award from the Swedish Foundation for Strategic Research in 2007, and a Consolidator Grant from the Swedish Research Council in 2016. He has served on the editorial boards of IEEE Transactions on Automatic Control and the IFAC Journal Automatica.
\end{IEEEbiography}

\begin{IEEEbiography}[{\includegraphics[width=1in,height=1.25in,clip]{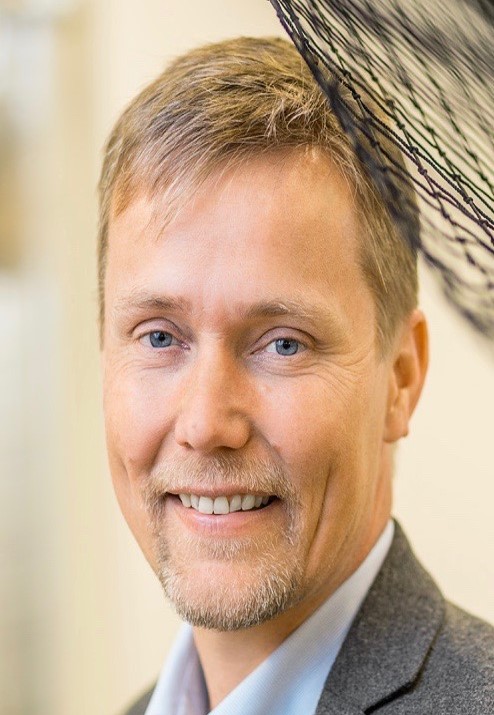}}]{Karl H. Johansson}
is Director of Digital Futures adn Professor with the School of Electrical Engineering and Computer Science at KTH Royal Institute of Technology in Sweden. He received MSc degree in Electrical Engineering and PhD in Automatic Control from Lund University. He has held visiting positions at UC Berkeley, Caltech, NTU, HKUST Institute of Advanced Studies, and NTNU. His research interests are in networked control systems and cyber-physical systems with applications in transportation, energy, and automation networks. He is President of the European Control Association and member of the IFAC Council, and has served on the IEEE Control Systems Society Board of Governors and the Swedish Scientific Council for Natural Sciences and Engineering Sciences. He has received several best paper awards and other distinctions from IEEE, IFAC, and ACM. He has been awarded Swedish Research Council Distinguished Professor, Wallenberg Scholar with the Knut and Alice Wallenberg Foundation, Future Research Leader Award from the Swedish Foundation for Strategic Research, the triennial IFAC Young Author Prize, and IEEE Control Systems Society Distinguished Lecturer. He is Fellow of the IEEE and the Royal Swedish Academy of Engineering Sciences.
\end{IEEEbiography}


\begin{thebibliography}{99}

\bibitem{Sandberg15}
	H.~Sandberg, S.~Amin, and K.\,H.~Johansson, ``Cyberphysical security in networked control systems,'' {\it IEEE Control Systems Magazine},  vol.~35, No.~1, pp.~20--23, 2015.
	
	\bibitem{Teixeira15}
	A.~Teixeira, I.~Shames, H.~Sandberg, and K.~H. Johansson, ``A secure control
	  framework for resource-limited adversaries,'' \emph{Automatica}, vol.~51, pp.~135--148, 2015.
	
	\bibitem{Amin09}
	S.~Amin, A.~A.~C{\'a}rdenas, and S.~S.~Sastry, ``Safe and secure networked
	control systems under denial-of-service attacks,'' in {\it Hybrid Systems:
		Computation and Control}, 2009, pp.~31--45.
		
		\bibitem{Sundaram11}
		S.~Sundaram and C.\,N.~Hadjicostis,
		``Distributed function calculation via linear iterative strategies in the presence of malicious agents,'' {\it IEEE Transactions on Automatic Control}, vol.~56, no.~7, pp.~1495--1508, 2011.
		
		\bibitem{Basar15}
		Q.~Zhu and T.~Basar,
		``Game-theoretic methods for robustness, security, and resilience of cyber physical control systems: games-in-games principle for optimal cross-layer resilient control systems,''
		{\it IEEE Control Systems Magazine}, vol.~35, no.~1, pp.~46--65, 2015.
		
		
		
		\bibitem{Langner11}
			R.~Langner, ``Stuxnet: Dissecting a cyberwarfare weapon," {\em IEEE Security~Privacy}, vol.~9, no.~3, pp.~49--51, 2011.
			
		
			
			\bibitem{Slay07}
			J.~Slay and M.~Miller, {\it Lessons Learned from the Maroochy Water Breach}, Springer, 2007.


		

	
	\bibitem{Pasqualetti13TAC}
	F.~Pasqualetti, F.~D\"{o}rfler, and F.~Bullo, ``Attack detection and identification in cyber-physical systems,'' {\it IEEE Transactions on Automatic Control}, vol.~58, no.~11, pp.~2715--2729, 2013.
	
	\bibitem{Pasqualetti15}
	F.~Pasqualetti, F.~D\"{o}rfler, and F.~Bullo, ``A divide-and-conquer approach to
	distributed attack identification,'' in \emph{Proceedings of the 54th IEEE
	Conference on Decision and Control}, 2015, pp.~5802--5807.
	
	\bibitem{Fawzi14}
H.~Fawzi, P.~Tabuada, and S.~Diggavi, ``Secure estimation and control for cyber-physical systems under adversarial attacks,'' {\it IEEE Trans.~Autom.~Control}, vol.~59, no.~6, pp.~1454--1467, 2014.

	\bibitem{Chong15ACC}
	M.\,S.~Chong, M.~Wakaiki, and J.\,P.~Hespanha, ``Observability of linear systems under adversarial attacks,'' in {\it Proc.~American Control Conference}, 2015, pp.~2439--2444.
								
\bibitem{Shoukry16TAC}
Y.~Shoukry and P.~Tabuada, ``Event-triggered state observers for sparse sensor
noise/attacks,'' \emph{IEEE Transactions on Automatic Control}, vol.~61,
no.~8, pp.~2079--2091, 2016.
		
		
		\bibitem{Chanhwa19TAC}
		C.~Lee, H.~Shim, and Y.~Eun, ``On redundant observability: from security index
		to attack detection and resilient state estimation,'' \emph{IEEE
Transactions on Automatic Control}, vol.~64, no.~2, pp.~775--782, 2019.
		
		\bibitem{jeon2016resilient}
		H.~Jeon, S.~Aum, H.~Shim, and Y.~Eun, ``Resilient state estimation for control
		systems using multiple observers and median operation,'' \emph{Mathematical
		Problems in Engineering}, vol.~2016, 2016.
		
	
	\bibitem{Jeong21TAC}
		Y.~Jeong and Y.~Eun,
		``A robust and resilient state estimation for linear systems,''
		{\it IEEE Transactions on Automatic Control},
		vol.~67, no.~5, pp.~2626--2632, 2022.
		
		
	   
\bibitem{Mitra16CDC}
  A.~Mitra and S.~Sundaram, ``Secure distributed observers for a class of linear
    time invariant systems in the presence of {B}yzantine adversaries,'' in
    \emph{Proceedings of the 55th IEEE Conference on Decision and Control}, 2016,
    pp. 2709--2714.
	      
\bibitem{Junsoo18}
J.~Kim, J.~G. Lee, C.~Lee, H.~Shim, and J.~H. Seo, ``Local identification of
  sensor attack and distributed resilient state estimation for linear
  systems,'' in \emph{Proceedings of the 57th IEEE Conference on Decision and
  Control}, 2018, pp.~2056--2061.
  
   \bibitem{Chen18TAC}
        Y.~Chen, S.~Kar, and J.\,M.\,F.~Moura, ``Resilient distributed estimation: sensor attacks,'' \emph{IEEE Transactions on Automatic Control}, vol.~64, no.~9, pp.~3772--3779, 2019.
        
        \bibitem{Lee20TAC}
        J.\,G.~Lee, J.~Kim, and H.~Shim, ``Fully distributed resilient state estimation based on distributed median solver,'' \emph{IEEE Transactions on Automatic Control}, vol.~65, no.~9, pp.~3935--3942, 2020.
  

      
      \bibitem{Kim19TAC}
      	J.~Kim, C.~Lee, H.~Shim, Y.~Eun, and J.\,H.~Seo,
      	``Detection of sensor attack and resilient state estimation for uniformly observable nonlinear systems having redundant sensors,''
      	{\it IEEE Transactions on Automatic Control}, vol.~64, no.~3, pp.~1162--1169, 2019.
      	ArXiv:1805.07944 [cs.SY].
      	
      	\bibitem{Chong20CDC}
      	M.\,S.~Chong, H.~Sandberg, and J.\,P.~Hespanha, ``A secure state estimation algorithm for nonlinear systems under sensor attacks,'' in {\it Proc.~59th IEEE Conference on Decision and Control}, 2020, pp.~5743--5748.
      	
      	\bibitem{Shou15CDC}
      	Y.~Shoukry, P.~Nuzzo, N.~Bezzo, A.\,L.~Sangiovanni-Vincentelli, S.\,A.~Seshiz, and P.~Tabuada, ``Secure state reconstruction in differentially flat systems under sensor attacks using satisfiability modulo theory solving,'' in {\it Proc.~54th IEEE Conference on Decision Control}, 2015, pp.~3804--3809.
      	
      	\bibitem{Mao22TAC}
      	Y.~Mao, A.~Mitra, S.~Sundaram, and P.~Tabuada, ``On the computational complexity of the secure state-reconstruction
      	problem,'' \emph{Automatica}, vol.~136, pp.~110083, 2022.
      	
      		\bibitem{Liu11}
      				Y.~Liu, P.~Ning, and M.\,K.~Reiter, ``False data injection attacks against state estimation in electric power grids," {\it ACM Trans.~Inform.~Syst.~Security}, vol.~14, no.~1, pp.~13:1--13:33, 2011.
      				
      				\bibitem{shoukry13springer}
      				Y.~Shoukry, P.~Martin, P.~Tabuada, and M.~Srivastava, ``Non-invasive spoofing attacks for anti-lock braking systems,''
      				{\it International Conference on Cryptographic Hardware and Embedded Systems}, 2013, pp.~55--72.
      	
      	\bibitem{Frank90}
      		P.\,M.~Frank, ``Fault diagnosis in dynamic systems using analytical and knowledge-based redundancy: A survey and some new results,'' {\it Automatica}, vol.~26, no.~3, pp.~459--474, 1990.
      				      	
      		\bibitem{Ding08}
      		S.\,X.~Ding, {\it Model-based Fault Diagnosis Techniques: Design Schemes, Algorithms, and Tools}, Springer Science \& Business Medea, 2008.
      		
      		\bibitem{Tao05}
      		E.\,J.~Candes and T.~Tao, ``Decoding by linear programming," {\it IEEE Transactions on Information Theory}, vol.~51, no.~12, pp.~4203--4215, 2005.
      		      	      		
      		\bibitem{Donoho06}
      		D.\,L.~Donoho, ``Compressed sensing," {\it IEEE Transactions on Information Theory}, vol.~52, no.~4, pp.~1289--1306, 2006.
      	
      	\bibitem{Gauthier92TAC}
      			J.\,P.~Gauthier, H.~Hammouri, and S.~Othman, ``A simple observer for nonlinear systems: applications to bioreactors,'' {\it IEEE Transactions on Automatic Control}, vol.~37, no.~6, pp.~875--880, 1992.
      	
      		\bibitem{Schwartz69}
      			J.\,T.~Schwartz, {\it Nonlinear Functional Analysis}, New York: Gordon and Breach Science, 1969.
      			


\end{thebibliography}
\end{document}